\documentclass[aps,prb,amsmath,amssymb,twocolumn,showpacs]{revtex4}
\usepackage{graphicx}
\usepackage{bm}
\usepackage{bbold}

\def\br{ \bm{r} }
\def\bk{ \bm{k} }

\def\bgam{ \bm{\gamma} }
\def\im{ \mathrm{Im}\, }
\def\re{ \mathrm{Re}\, }
\def\sgn{\, \mathrm{sgn}\, }

\def\Tr{\,\mathrm{Tr}\,}

\begin{document}
\title{Fermionic boundary modes in two-dimensional noncentrosymmetric superconductors}

\author{K. V. Samokhin and S. P. Mukherjee}

\affiliation{Department of Physics, Brock University, St. Catharines, Ontario L2S 3A1, Canada}
\date{\today}

\begin{abstract}
We calculate the spectrum of the Andreev boundary modes in a two-dimensional superconductor formed at an interface between two different non-superconducting materials, e.g. insulating oxides. Inversion symmetry is absent in this system, and
both the electron band structure and the superconducting pairing are strongly affected by the spin-orbit coupling of the Rashba type.  
We consider isotropic $s$-wave pairing states, both with and without time-reversal symmetry breaking, as well as various $d$-wave states. In all cases, there exist subgap Andreev boundary states, whose properties, 
in particular, the number and location of the zero-energy modes, qualitatively depend on the gap symmetry and the spin-orbit coupling strength.  
\end{abstract}

\pacs{74.20.-z}

\maketitle

\section{Introduction}
\label{sec: introduction}

Superconducting materials without inversion symmetry have recently become a subject of rapidly growing interest, see Refs. \onlinecite{NCSC-book} and \onlinecite{Kneidinger-exp-review} for a review and references. 
Due to the qualitative changes in their band structure caused by the spin-orbit (SO) coupling of electrons with the crystal lattice, properties of these materials differ significantly from the 
predictions of the standard Bardeen-Cooper-Schrieffer (BCS) theory of superconductivity. In a nutshell, the SO coupling in a noncentrosymmetric crystal lifts the spin degeneracy of the Bloch states almost everywhere in the Brillouin zone (BZ).
Spin is no longer a good quantum number, and the nondegenerate bands are instead labelled by ``helicity'' and have a nontrivial topology in the momentum space. 
If the SO band splitting is large compared to all superconducting energy scales (which is the case in real materials), then the Cooper pairing occurs only between time-reversed quasiparticle states of the same helicity, 
with profound consequences for superconductivity.\cite{MinSig-chapter} 

While noncentrosymmetric superconductivity has mostly been observed in three-dimensional materials, it can also be realized in two dimensions (2D), for example, at the interface LAO/STO between two band insulators, LaAlO$_3$ and SrTiO$_3$ 
(Ref. \onlinecite{LAO-STO}). Other similar systems include the LSCO/LCO or LTO/STO interfaces between various metallic or insulating oxides, and also surfaces of doped insulators, such as STO and possibly WO$_3$, see 
Refs. \onlinecite{interface-SC-1} and \onlinecite{interface-SC-2} for a review. The superconducting critical temperature $T_c$ can be as high as 109 K, for FeSe monolayers deposited on doped STO substrates.\cite{FeSe-layers} 
In all these systems the inversion symmetry is broken due to the different nature of the materials sandwiching the conducting layer. As an added bonus, the SO coupling strength in the oxide interfaces can be controlled 
by applying an external gate voltage. For instance, the SO band splitting in the 2D electron gas at the LAO/STO interface can be tuned between 1 and 10 meV, while the maximum value of $T_c$ is about 0.3 K (Ref. \onlinecite{SOC-2D}). 

The qualitative significance of the electron-lattice SO coupling makes noncentrosymmetric materials promising candidates for applications to spintronics,\cite{FY-chapter} as well as for topological superconductivity.\cite{SF09}
The hallmark property of topological superfluids and superconductors is that, while fermionic excitations in the bulk are gapped, there are zero-energy boundary modes propagating along the surface of the system, 
see Refs. \onlinecite{Volovik-book} and \onlinecite{top-SC}. These modes are topologically protected against sufficiently small perturbations, can carry charge and spin currents, and also lead to prominent peaks 
in the tunneling conductance.\cite{Hu94,ZBCP} 

In this paper we study the spectrum of the fermionic modes localized near the boundary of a semi-infinite 2D noncentrosymmetric superconductor. Previous works on this subject have focused mostly 
on time-reversal (TR) invariant isotropic pairing states, see Ref. \onlinecite{EIT-chapter} for a review.
The Bogoliubov-de Gennes (BdG) equation in a half-plane with the SO coupling of the Rashba form,\cite{Rashba-model} was solved in Ref. \onlinecite{TYBN09}, 
while a different approach, based on the Eilenberger equations without the SO band splitting, corresponding to a weak SO coupling limit, was developed in Refs. \onlinecite{Inio07} and \onlinecite{VVE08}. 
The main result is that the fermionic boundary modes are present only if the ``protected'' spin-triplet component\cite{FAKS04} of the gap function is greater than the spin-singlet component, which puts the system in 
a $Z_2$-nontrivial topological class.\cite{SRFL08} 
The effects of the TR symmetry breaking by an external magnetic field have been studied in Ref. \onlinecite{STF09}, where the Zeeman interaction was included in the singlet-triplet-mixing BdG Hamiltonian. 
It was found that the gapless boundary modes can appear even in the absence of the triplet component, if the field is sufficiently strong.

Our goal is twofold. First, we would like to fill the gaps in the literature and study the boundary mode spectra in (i) a general isotropic superconducting state, in which the TR symmetry is broken intrinsically, i.e. without any external field 
(according to the symmetry classification of the stable states in 2D noncentrosymmetric superconductors,\cite{Sam15-1} such states are possible on phenomenological grounds), and (ii) anisotropically paired states, both with and without gap nodes 
and/or TR symmetry breaking. Since the boundary modes can be probed in tunneling experiments, understanding their spectra can help determine the pairing symmetry in the bulk. Second, we aim to go beyond the weak SO coupling limit and develop a formalism which is applicable for any SO coupling strength, pairing symmetry, and potentially any type of the surface scattering.    

The treatment of the fermionic boundary modes in this paper is based on the semiclassical, or Andreev, equations\cite{Andreev-approx} for the quasiparticle wave functions in the helicity representation. 
The standard theoretical approach,\cite{Book} which describes superconductivity in terms of spin-singlet and spin-triplet gap functions, is not justified in noncentrosymmetric materials with a large SO splitting of nondegenerate bands.
Instead, one should work in the helicity representation and construct the pairing interaction using the exact band eigenstates, which incorporate all effects of the noncentrosymmetric lattice potential and the strong SO coupling. 
In the semiclassical picture, the Fermi-surface quasiparticles of definite helicity propagate along straight lines in the bulk, while the surface scattering is described
by an effective boundary condition formulated in terms of the surface $S$-matrix mixing the Andreev amplitudes for different semiclassical trajectories.\cite{Shel-bc} 

The paper is organized as follows. In Sec. \ref{sec: general}, we introduce the helicity representation, using the Rashba model as an example, and discuss the peculiarities of the superconducting pairing in the nondegenerate helicity bands.
In Sec. \ref{sec: ABS}, we derive general equations for the energy of the fermionic boundary modes as a function of the momentum along the surface. The results strongly depend on the number of the surface 
scattering channels. In Secs. \ref{sec: isotropic} and \ref{sec: chiral d-wave}, the boundary mode spectrum is calculated in the fully gapped states, $s$-wave and the chiral $d$-wave, respectively, with particular attention given 
to the fate of the zero-energy modes. In Sec. \ref{sec: d-wave}, the TR invariant (nonchiral) $d$-wave states with gap nodes are examined. Sec. \ref{sec: Conclusions} concludes with a summary of our results.
Throughout the paper we use the units in which $\hbar=k_B=1$, neglecting, in particular, the difference between the quasiparticle momentum and wavevector.

\section{Superconductivity in nondegenerate bands}
\label{sec: general}

The minimal model that captures the essential physics of a 2D electron gas with an asymmetric SO coupling is described by the following Hamiltonian: 
\begin{equation}
\label{H-general}
    \hat H_0=\sum\limits_{\bk,\alpha\beta}h_{\alpha\beta}(\bk)\hat a^\dagger_{\bk\alpha}\hat a_{\bk\beta}.
\end{equation}
Here $\bk=(k_x,k_y)$ is a 2D wavevector, $\alpha,\beta=\uparrow,\downarrow$ are spin indices, $\hat h(\bk)=\epsilon(\bk)\hat\sigma_0+\bgam(\bk)\hat{\bm{\sigma}}$, and $\hat{\bm{\sigma}}$ are the Pauli matrices. 
The first term in $\hat h$ is the ``bare'' band dispersion without the SO coupling. The chemical potential, which is assumed to be equal to the Fermi energy $\epsilon_F$, is included in $\epsilon(\bk)$. 
The second term describes the SO coupling of 2D electrons with their noncentrosymmetric environment. For example, at an interface between two insulating oxides, this SO coupling is due to the intrinsic electric field normal to
the interface, which compensates the charge discontinuity between the two sides.\cite{interface-SC-2} 
Due to the TR invariance of the normal state, we have $\epsilon(\bk)=\epsilon(-\bk)$ and $\bgam(\bk)=-\bgam(-\bk)$, with additional contraints imposed by the 2D point group symmetry.\cite{Sam15-1}
Diagonalization of the Hamiltonian (\ref{H-general}) produces two nondegenerate bands
\begin{equation}
\label{xis-general}
  \xi_\lambda(\bk)=\epsilon(\bk)+\lambda|\bgam(\bk)|=\xi_\lambda(-\bk),
\end{equation}
labelled by helicity $\lambda=\pm$. Physically, the helicity corresponds to the spin projection on the direction of the SO coupling $\bgam(\bk)$.

The superconducting pairing takes places between the time-reversed Bloch states of the same helicity, $|\bk,\lambda\rangle$ and $K|\bk,\lambda\rangle$, which belong to $\bk$ and $-\bk$, respectively,
and have the same energy. Recall that the TR operator for spin-1/2 particles has the form $K=i\hat\sigma_2K_0$, where $K_0$ is complex conjugation. 
Since the bands are nondegenerate almost everywhere in the BZ, one has 
\begin{equation}
\label{t-lambda}
  K|\bk,\lambda\rangle=t_\lambda(\bk)|-\bk,\lambda\rangle,
\end{equation}
where $t_\lambda(\bk)=-t_\lambda(-\bk)$ is a phase factor,\cite{t-factor} which cannot be removed by a gauge transformation of the Bloch states. While the phase factor is not defined at the band degeneracy points, one can 
use its winding numbers around these points to introduce a $Z_2$ topological invariant of the normal-state band structure.\cite{Sam15-1}

To make analytical progress, we will use the isotropic effective mass approximation for the bare band dispersion and a particular form of the SO coupling known as the Rashba model, see Ref. \onlinecite{Rashba-model} 
and the references therein, which is described by the following Hamiltonian:
\begin{equation}
\label{H-isotropic-Rashba}
  \hat h(\bk)=\left(\frac{\bk^2}{2m^*}-\epsilon_F\right)\hat\sigma_0+\gamma_0(k_y\hat\sigma_x-k_x\hat\sigma_y),
\end{equation}
where $\epsilon_F=k_F^2/2m^*$ and $k_F$ is the Fermi wave vector in the absence of the SO coupling.
For the helicity bands we obtain
\begin{equation}
\label{xis-Rashba}
    \xi_\lambda(\bk)=\frac{|\bk|^2-k_F^2}{2m^*}+\lambda\gamma_0|\bk|,
\end{equation}
assuming $\gamma_0>0$. Although the two Fermi surfaces have different radii: 
$$
  k_{F,\lambda}=\sqrt{k_F^2+(m^*\gamma_0)^2}-\lambda m^*\gamma_0, 
$$
i.e. $k_{F,-}>k_{F,+}$, the Fermi velocities are the same in both bands:
\begin{equation}
\label{v-F-pm}
  \bm{v}_{F,\lambda}=v_F\frac{\bk}{|\bk|},\quad v_F=\frac{1}{m^*}\sqrt{k_F^2+(m^*\gamma_0)^2}.
\end{equation}
It is convenient to introduce the parameter
\begin{equation}
\label{rho-def}
  \rho=\frac{k_{F,+}}{k_{F,-}},\quad 0<\rho\leq 1,
\end{equation}
as a dimensionless measure of the SO coupling strength. Zero SO coupling corresponds to $\rho=1$, while in the limit of very strong SO coupling, we have $\rho\to 0$ and the minority ($\lambda=+$) Fermi surface shrinks to a point. 
The eigenstates of the Rashba Hamiltonian (\ref{H-isotropic-Rashba}) have the form
\begin{equation}
\label{Rashba-eigenstates}
  \chi_\lambda(\bk)=\frac{1}{\sqrt{2}}\left(\begin{array}{c}
                                       1 \\ -i\lambda e^{i\varphi_{\bk}}
                                      \end{array}\right).
\end{equation}
where $\varphi_{\bk}=\tan^{-1}(k_y/k_x)$ is the angle between $\bk$ and the positive $x$ axis.
It follows from Eqs. (\ref{t-lambda}) and (\ref{Rashba-eigenstates}) that the phase factor connecting the time-reversed Rashba eigenstates is given by $t_\lambda(\bk)=i\lambda e^{-i\varphi_{\bk}}$.

We now use the basis of the exact helicity states $|\bk,\lambda\rangle$ to construct the pairing interaction between electrons.
Assuming a BCS-like mechanism of superconductivity, this interaction is only effective near the 2D Fermi surface. The latter is defined, in the $\lambda$th band, by the equation $\xi_\lambda(\bk)=0$.
In real materials the energy scales associated with superconductivity, including the critical temperature $T_c$ and the BCS energy cutoff, are much smaller than the SO band splitting $E_{SO}$ (in the Rashba model, $E_{SO}=2\gamma_0k_F$). 
This means that the Fermi surfaces are sufficiently well separated to suppress the pairing of electrons with opposite helicities, which leads to the following mean-field Hamiltonian:
\begin{eqnarray}
\label{H_mean-field}
     \hat H &=& \sum_{\bk,\lambda=\pm}\xi_\lambda(\bk)\hat c^\dagger_{\bk,\lambda}\hat c_{\bk,\lambda}\nonumber\\
     &&+\frac{1}{2}\sum_{\bk,\lambda=\pm}\left[\tilde\Delta_\lambda(\bk)\hat c^\dagger_{\bk,\lambda}\hat{\tilde c}^\dagger_{\bk,\lambda}+\tilde\Delta^*_\lambda(\bk)\hat{\tilde c}_{\bk,\lambda}\hat c_{\bk,\lambda}\right].
\end{eqnarray}
The Cooper pairing takes place between the states $|\bk,\lambda\rangle$ and $K|\bk,\lambda\rangle$, and $\hat{\tilde c}^\dagger_{\bk,\lambda}\equiv K\hat c^\dagger_{\bk,\lambda}K^{-1}=t_\lambda(\bk)\hat c^\dagger_{-\bk,\lambda}$.
Due to the anticommutation of the fermion creation and annihilation operators, the gap functions in the helicity representation are even in $\bk$: 
\begin{equation}
\label{Delta-even}
  \tilde\Delta_\lambda(\bk)=\tilde\Delta_\lambda(-\bk).
\end{equation}
The momentum dependence of the gap functions, in particular the presence and location of the gap nodes, is determined by the irreducible representations of the 2D point group, see Ref. \onlinecite{Sam15-1} for a detailed analysis. 
In this paper we focus on the $s$-wave and $d$-wave pairing states, corresponding to the two lowest possible values of the pair angular momentum compatible with the condition (\ref{Delta-even}). 

The model defined by the Hamiltonian (\ref{H_mean-field}) is formally similar to the two-band BCS theory, which has been recently applied to MgB$_2$, iron-based high-temperature superconductors, and other materials.\cite{two-band} 
Note though that in our case, the bands are nondegenerate and the pairing symmetry classification is different, see Ref. \onlinecite{Sam15-1}.
In general, the number of bands split by the SO coupling can be greater than two, leading to multi-component superconducting order parameters and complex phase diagrams. It has been shown\cite{TRB-states} that some of the stable states 
found by minimizing the Ginzburg-Landau free energy with two or more bands break the TR symmetry. 
We consider the TR symmetry-breaking $s$-wave and $d$-wave states in Secs. \ref{sec: isotropic} and \ref{sec: chiral d-wave}, respectively. 

To conclude this section, we note that the helicity band description of noncentrosymmetric superconductivity with a strong SO coupling can be easily translated into the language of spin-singlet and spin-triplet components. The gap function 
in the spin representation contains both the singlet and triplet parts, given by $\psi\sim\tilde\Delta_++\tilde\Delta_-$ and $\bm{d}\sim(\tilde\Delta_+-\tilde\Delta_-)\hat{\bgam}$, respectively.\cite{NCSC-book}
In the limit of a local BCS attractive interaction, both gap functions are the same: $\tilde\Delta_+=\tilde\Delta_-=\Delta_0$, which corresponds to a purely singlet isotropic pairing, regardless of the SO coupling strength.
Any difference between $\tilde\Delta_+$ and $\tilde\Delta_-$, giving rise to the ``protected'' triplet order parameter $\bm{d}(\bk)\parallel\bgam(\bk)$ (Ref. \onlinecite{FAKS04}), 
is only possible if the pairing interaction contains a triplet component.

\section{Fermionic boundary modes}
\label{sec: ABS}

Consider a 2D noncentrosymmetric superconductor occupying the positive-$x$ half-plane, in which quasiparticles are reflected specularly from an atomically smooth straight boundary at $x=0$. 
To make analytical progress, we neglect self-consistency and assume that the order parameter is uniform. Translational invariance along the boundary implies that $k_y$ is a good quantum number. 
Then, the Bogoliubov quasiparticle wave function in each band is a two-component (electron-hole) spinor, which can be represented in the semiclassical, or Andreev, approximation\cite{Andreev-approx} as $e^{i\bk_{\lambda,n}\br}\psi_{\lambda,n}(x)$, 
where $\bk_{\lambda,n}$ is a Fermi-surface wavevector in the $\lambda$th band and $n$ labels the roots of the equation
\begin{equation}
\label{FS-intersect-eq}
	\xi_\lambda(\bk)=0
\end{equation} 
at given $k_y$. The helicity band dispersions for a general antisymmetric SO coupling are given by Eq. (\ref{xis-general}). 
The Andreev envelope function $\psi_{\lambda,n}$ varies slowly on the scale of the Fermi wavelength and satisfies the following equation:    
\begin{equation}
\label{And-eq-gen}
	\left(\begin{array}{cc}
		-iv_{\lambda,n}\nabla_x & \Delta_{\lambda,n} \\
		\Delta^*_{\lambda,n} & iv_{\lambda,n}\nabla_x
	\end{array}\right)\psi=E\psi.
\end{equation}
Here $v_{\lambda,n}=(\partial\xi_\lambda/\partial k_x)|_{\bk=\bk_{\lambda,n}}$ is the $x$-projection of the Fermi velocity 
and 
$$
  \Delta_{\lambda,n}\equiv\Delta(\bk_{\lambda,n})=\tilde\Delta_\lambda(\bk_{\lambda,n})
$$ 
is a shorthand notation for the gap function sensed by the quasiparticles in the $\lambda$th band propagating with the wavevector $\bk_{\lambda,n}$.

At given momentum along the surface, Eq. (\ref{FS-intersect-eq}) can have several solutions, determined by the band structure. 
Depending on the direction of propagation, the corresponding Andreev states are classified as either incident, for which $v_{\lambda,n}<0$, or reflected, for which $v_{\lambda,n}>0$. 
For $v_{\lambda,n}=0$, the quasiparticles move along the surface and the semiclassical approximation is not applicable.  

We focus on the quasiparticle states localized near the surface, which are called the Andreev bound states (ABSs). The corresponding solution of Eq. (\ref{And-eq-gen}) has the form 
$\psi_{\lambda,n}(x)=\phi(\bk_{\lambda,n})e^{-\Omega_{\lambda,n}x/|v_{\lambda,n}|}$, where 
\begin{eqnarray}
\label{Andreev amplitude}
	\phi(\bk_{\lambda,n}) &\equiv& \psi_{\lambda,n}(x=0)\nonumber\\
	&=& C(\bk_{\lambda,n})\left(\begin{array}{c}
		\dfrac{\Delta_{\lambda,n}}{E-i\Omega_{\lambda,n}\sgn v_{\lambda,n}} \\ 1
	\end{array}\right),
\end{eqnarray}
$\Omega_{\lambda,n}=\sqrt{|\Delta_{\lambda,n}|^2-E^2}$, and $C(\bk_{\lambda,n})$ is a coefficient. 
The semiclassical approximation breaks down near the surface due to the rapid variation of the lattice potential, which causes elastic transitions between the states corresponding to different Fermi wavevectors, in particular, between the 
states of different helicity. Therefore, the ABS wave function away from the surface becomes a superposition of the solutions corresponding to all possible Fermi wavevectors $\bk_{\lambda,n}$ at given $k_y$:
\begin{equation}
\label{Psi-ABS}
  \Psi_{k_y}(\br)=\sum_{\lambda,n}\phi(\bk_{\lambda,n})e^{i\bk_{\lambda,n}\br}e^{-\Omega_{\lambda,n} x/|v_{\lambda,n}|}.
\end{equation}
In order for the wave function to be localized near the surface, the energy has to be inside the bulk gaps, i.e. $|E|<|\Delta_{\lambda,n}|$ for all $\bk_{\lambda,n}$.

Suppose that at given momentum along the surface the total number of roots of the equations (\ref{FS-intersect-eq}) in both helicity bands is equal to $2N$, describing the incident and reflected Fermi wavevectors 
$\bk^{\mathrm{in}}_{1},...,\bk^{\mathrm{in}}_{N}$ and $\bk^{\mathrm{out}}_{1},...,\bk^{\mathrm{out}}_{N}$, respectively. Following Ref. \onlinecite{Shel-bc}, we describe the surface scattering by an effective boundary condition, 
which expresses the Andreev amplitudes at $x=0$ for the reflected waves in terms of those for the incident waves as follows:
\begin{equation}
\label{Shelankov-bc}
  \phi(\bk^{\mathrm{out}}_{i})=\sum_{j=1}^{N} S_{ij}\phi(\bk^{\mathrm{in}}_{j}).
\end{equation}
Here $\hat S$ is an $N\times N$ unitary matrix and $i,j=1,...,N$ label the surface scattering channels. The $S$-matrix is an electron-hole scalar, which is determined by the microscopic details in the normal state.

\begin{figure}
\includegraphics[width=5.8cm]{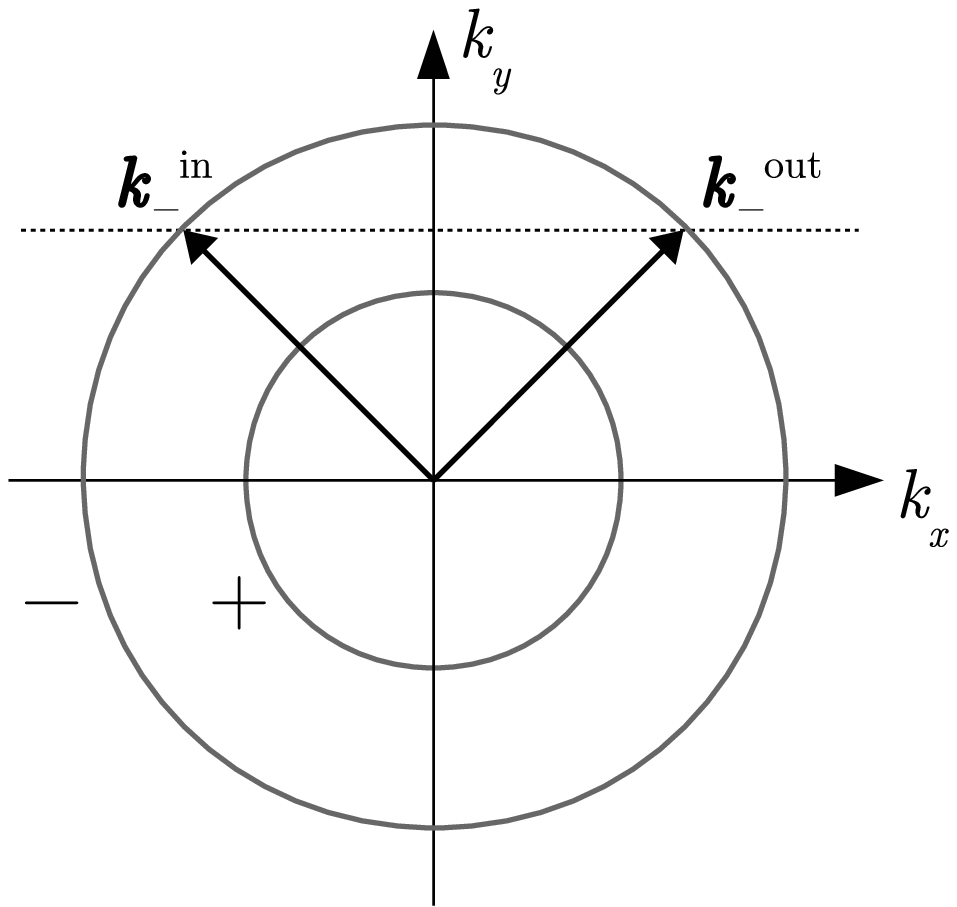}
\caption{The incident and reflected wavevectors for $N=1$ ($k_{F,+}<|k_y|<k_{F,-}$). The circular Fermi surfaces with $\lambda=-$ and $+$ correspond to the majority and minority helicity 
bands in the Rashba model.}
\label{fig: Rashba-isotropic-N-1}
\end{figure}

\begin{figure}
\includegraphics[width=5.8cm]{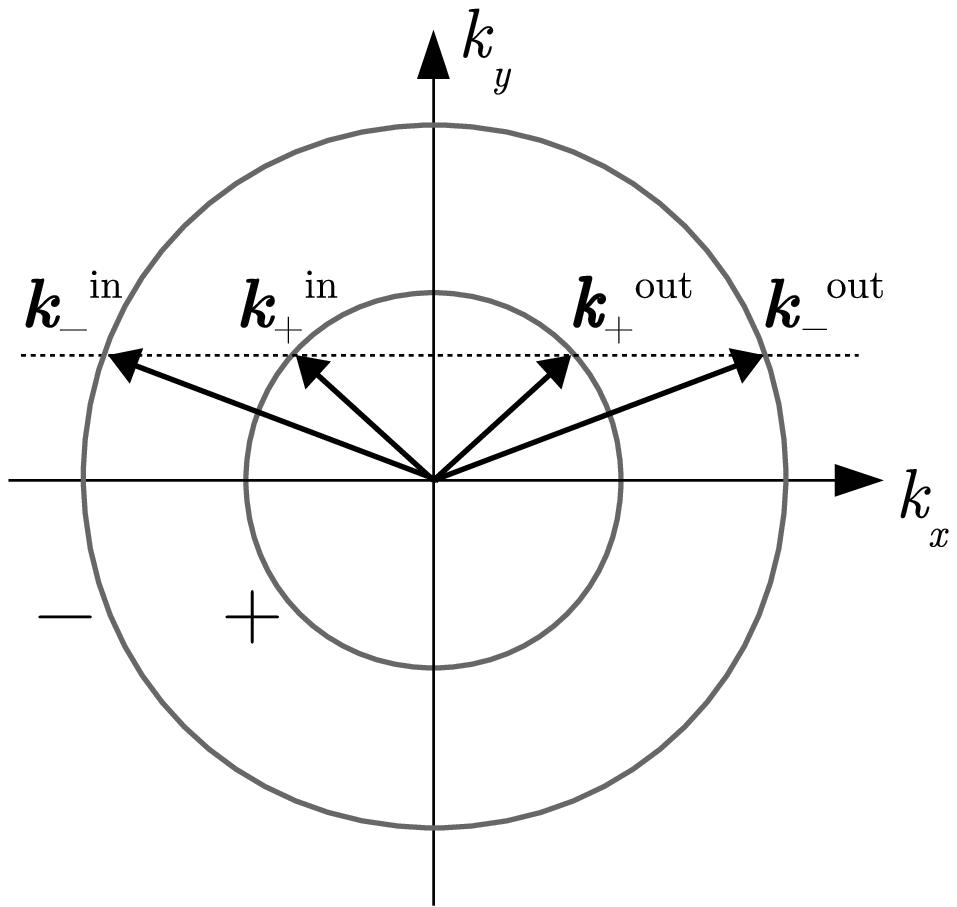}
\caption{The incident and reflected wavevectors for $N=2$ ($|k_y|<k_{F,+}$). The circular Fermi surfaces with $\lambda=-$ and $+$ correspond to the majority and minority helicity
bands in the Rashba model.}
\label{fig: Rashba-isotropic-N-2}
\end{figure}

According to Eq. (\ref{Andreev amplitude}), the Andreev amplitudes for the incident waves have the form 
\begin{equation}
\label{phi-in}
  \phi(\bk^{\mathrm{in}}_{i})=C(\bk^{\mathrm{in}}_{i})\left(\begin{array}{c}
		\alpha^{\mathrm{in}}_i \\ 1
	\end{array}\right)
\end{equation}
where
\begin{equation}
\label{alpha-in-def}
  \alpha^{\mathrm{in}}_i=\frac{\Delta(\bk^{\mathrm{in}}_{i})}{E+i\sqrt{|\Delta(\bk^{\mathrm{in}}_{i})|^2-E^2}}.
\end{equation}
For the reflected wave amplitudes we have 
\begin{equation}
\label{phi-out}
  \phi(\bk^{\mathrm{out}}_{i})=C(\bk^{\mathrm{out}}_{i})\left(\begin{array}{c}
		\alpha^{\mathrm{out}}_i \\ 1
	\end{array}\right),
\end{equation}
where
\begin{equation}
\label{alpha-out-def}
  \alpha^{\mathrm{out}}_i=\frac{\Delta(\bk^{\mathrm{out}}_{i})}{E-i\sqrt{|\Delta(\bk^{\mathrm{out}}_{i})|^2-E^2}}.
\end{equation}
Inserting Eqs. (\ref{phi-in}) and (\ref{phi-out}) into the boundary conditions (\ref{Shelankov-bc}), we obtain a homogeneous system of $2N$ linear equations for the coefficients $C(\bk^{\mathrm{in}}_{1}),...,C(\bk^{\mathrm{in}}_{N})$
and $C(\bk^{\mathrm{out}}_{1}),...,C(\bk^{\mathrm{out}}_N)$. Equating its determinant to zero yields an equation for the ABS energy $E(k_y)$. Below we consider two cases which can be treated analytically: one scattering channel in the majority 
($\lambda=-$) band, or two scattering channels, one in each band. These cases are illustrated in Figs. \ref{fig: Rashba-isotropic-N-1} and \ref{fig: Rashba-isotropic-N-2}, respectively, for the isotropic helicity bands in the Rashba model.

For $N=1$, the scattering matrix becomes just a single complex number (a pure phase). The energy equation then takes the simple form $\alpha^{\mathrm{in}}_-=\alpha^{\mathrm{out}}_-$, or  
\begin{equation}
\label{N-1-ABS-energy}
  \frac{E+i\sqrt{|\Delta(\bk^{\mathrm{in}}_{-})|^2-E^2}}{E-i\sqrt{|\Delta(\bk^{\mathrm{out}}_{-})|^2-E^2}}=\frac{\Delta(\bk^{\mathrm{in}}_{-})}{\Delta(\bk^{\mathrm{out}}_{-})},
\end{equation}
which remarkably does not contain any surface scattering details. It follows from this last equation that the subgap ABS can exist only if $\Delta(\bk^{\mathrm{out}}_{-})\neq\Delta(\bk^{\mathrm{in}}_{-})$, i.e. when the quasiparticles sense 
different gap functions before and after the surface reflection. This is similar to other systems in which the gap function variation along the quasiparticle's semiclassical trajectory leads to a bound state. Examples include the ABS near 
a surface of a $d$-wave or a chiral $p$-wave superconductor,\cite{Hu94,ZBCP} or near a superconducting domain wall.\cite{SC-DWs} 

For $N=2$, the boundary condition (\ref{Shelankov-bc}) takes the form
$$
  \left(\begin{array}{cccc}
  S_{--}\alpha^{\mathrm{in}}_- & S_{-+}\alpha^{\mathrm{in}}_+ & -\alpha^{\mathrm{out}}_- & 0 \\
  S_{--} & S_{-+} & -1 & 0 \\
  S_{+-}\alpha^{\mathrm{in}}_- & S_{++}\alpha^{\mathrm{in}}_+ & 0 & -\alpha^{\mathrm{out}}_+ \\
  S_{+-} & S_{++} & 0 & -1 
  \end{array}\right)
  \left(\begin{array}{c}
  C(\bk^{\mathrm{in}}_{-}) \\ C(\bk^{\mathrm{in}}_{+}) \\ C(\bk^{\mathrm{out}}_{-}) \\ C(\bk^{\mathrm{out}}_{+})
  \end{array}\right)=0.
$$
From this we obtain the following ABS energy equation at given $k_y$:
\begin{equation}
\label{N-2-ABS-energy}
  \frac{(\alpha^{\mathrm{in}}_--\alpha^{\mathrm{out}}_-)(\alpha^{\mathrm{in}}_+-\alpha^{\mathrm{out}}_+)}{(\alpha^{\mathrm{in}}_--\alpha^{\mathrm{out}}_+)(\alpha^{\mathrm{in}}_+-\alpha^{\mathrm{out}}_-)}=\frac{S_{-+}S_{+-}}{S_{--}S_{++}},
\end{equation}
where $\alpha^{\mathrm{in}}_\pm$ and $\alpha^{\mathrm{out}}_\pm$ are defined by Eqs. (\ref{alpha-in-def}) and (\ref{alpha-out-def}). Note that Eqs. (\ref{N-1-ABS-energy}) and (\ref{N-2-ABS-energy}) 
are valid for any gap symmetry and band structure, as long as the surface scattering is specular.

\section{S-wave pairing}
\label{sec: isotropic}

In this section we consider the pairing state described by the following gap functions:
\begin{equation}
\label{isotropic-Delta}
  \tilde\Delta_-(\bk)=\Delta_-,\quad \tilde\Delta_+(\bk)=\Delta_+e^{i\chi},
\end{equation}
where $\Delta_\pm\geq 0$ are the gap magnitudes. Due to the momentum-space isotropy it can be called the $s$-wave state. 
If the phase difference between the bands is equal to $0$ or $\pi$, as usually assumed, then the superconducting state is TR invariant. 
However, minimization of the phenomenological two-band Ginzburg-Landau theory can yield an arbitrary value of $\chi$, leading to the possibility of TR symmetry-breaking stable states. For this reason, we consider the general case
with $0\leq\chi\leq\pi$. While the quasiparticle spectrum in the bulk is fully gapped, there might exist the subgap surface states, whose energy depends on $\chi$. It is known\cite{TYBN09} that such states are present if $\chi=\pi$ 
(which corresponds to the dominant triplet component in the spin representation, see the end of Sec. \ref{sec: general}) and are absent if $\chi=0$. 

\begin{figure}
\includegraphics[width=4.5cm]{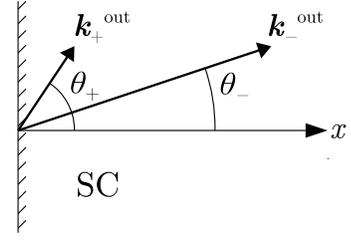}
\caption{The reflection angles at given $k_y$, for $N=2$.}
\label{fig: reflection angles}
\end{figure}

From this point on we focus on the isotropic Rashba model [Eq. (\ref{H-isotropic-Rashba})] in a half-plane, for which the $S$-matrix can be calculated explicitly, see Appendix \ref{app: S-matrix}. 
At given momentum along the surface, the directions of semiclassical trajectories can be characterized by the angles of reflection $\theta_-$ and $\theta_+$, as shown in Fig. \ref{fig: reflection angles}. We have
\begin{equation}
\label{ky-theta}
  k_y=k_{F,-}\sin\theta_-=k_{F,+}\sin\theta_+,
\end{equation}
so that $\bk^{\mathrm{in}}_{\lambda}=k_{F,\lambda}(-\cos\theta_\lambda,\sin\theta_\lambda)$ and $\bk^{\mathrm{out}}_{\lambda}=k_{F,\lambda}(\cos\theta_\lambda,\sin\theta_\lambda)$.

\subsection{$k_{F,+}<|k_y|<k_{F,-}$}
\label{sec: isotropic-N-1}

In this case, there is just one scattering channel, in the majority ($\lambda=-$) band, see Fig. \ref{fig: Rashba-isotropic-N-1}. The gap function $\tilde\Delta_-(\bk)$ has the same value $\Delta_-$ on the incident and reflected legs
of the semiclassical trajectory and the only solution of Eq. (\ref{N-1-ABS-energy}) is $|E(k_y)|=\Delta_-$. Therefore, there is no subgap ABS, regardless of the value of $\chi$.

\subsection{$|k_y|<k_{F,+}$}
\label{sec: isotropic-N-2}

This momentum range corresponds to $N=2$, see Fig. \ref{fig: Rashba-isotropic-N-2}, and the ABS energy as a function of $k_y$ is obtained by solving Eq. (\ref{N-2-ABS-energy}). The surface $S$-matrix for the isotropic Rashba model is given by
\begin{eqnarray}
\label{S-matrix-Rashba}
  && S_{--}=\frac{e^{-i\theta_-}-e^{i\theta_+}}{e^{i\theta_-}+e^{i\theta_+}},\nonumber\\
  && S_{-+}=S_{+-}=-\frac{2\sqrt{\cos\theta_-\cos\theta_+}}{e^{i\theta_-}+e^{i\theta_+}},\\
  && S_{++}=-\frac{e^{i\theta_-}-e^{-i\theta_+}}{e^{i\theta_-}+e^{i\theta_+}},\nonumber
\end{eqnarray}
see Eq. (\ref{S-matrix-N-2}). The ABS energy equation takes the form
\begin{equation}
\label{N-2-ABS-energy-Rashba}
  \frac{(\alpha^{\mathrm{in}}_--\alpha^{\mathrm{out}}_-)(\alpha^{\mathrm{in}}_+-\alpha^{\mathrm{out}}_+)}{(\alpha^{\mathrm{in}}_--\alpha^{\mathrm{out}}_+)(\alpha^{\mathrm{in}}_+-\alpha^{\mathrm{out}}_-)}=1-\frac{1}{\zeta},
\end{equation}
where
\begin{equation}
\label{zeta}
  \zeta(k_y)=\frac{1-\cos(\theta_-+\theta_+)}{1+\cos(\theta_--\theta_+)},\quad 0\leq\zeta\leq 1.
\end{equation}
Substituting here the gap functions (\ref{isotropic-Delta}), we arrive at the following equation for $E(k_y)$:
\begin{equation}
\label{ABS-energy-chi}
  \frac{E^2-\Delta_-\Delta_+\cos\chi}{\sqrt{(\Delta_-^2-E^2)(\Delta_+^2-E^2)}}=R, 
\end{equation}
where
\begin{equation}
\label{R-ky}
  R(k_y)=\frac{1+\zeta(k_y)}{1-\zeta(k_y)}=\frac{k_y^2+k_{F,-}k_{F,+}}{\sqrt{(k_{F,-}^2-k_y^2)(k_{F,+}^2-k_y^2)}}.
\end{equation}
The solution for the ABS energy has to be inside the bulk gaps, i.e. $|E|<\min(\Delta_-,\Delta_+)$. Since Eq. (\ref{ABS-energy-chi}) contains only $E^2$, there are two ABSs at each $k_y$, with energies $\pm|E(k_y)|$. 
Also, it is easy to see that the ABS spectrum is symmetric with respect to the inversion of the momentum parallel to the surface, i.e. $E(k_y)=E(-k_y)$.

\begin{figure}
\includegraphics[width=7cm]{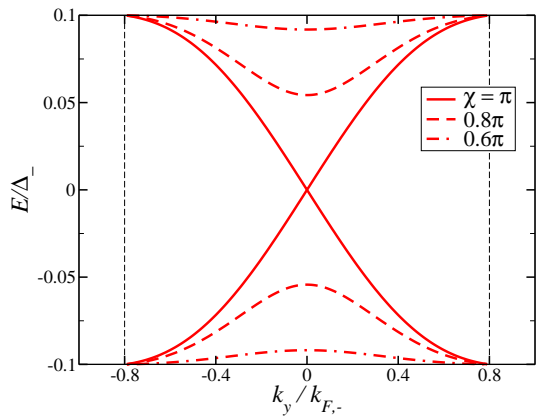}
\caption{(Color online) The surface ABS dispersion in the case of $s$-wave pairing, for $\Delta_+/\Delta_-=0.1$. The critical value of the phase difference is $\chi_c\simeq 0.47\pi$.}
\label{fig: gamma-0.1}
\end{figure}

\begin{figure}
\includegraphics[width=7cm]{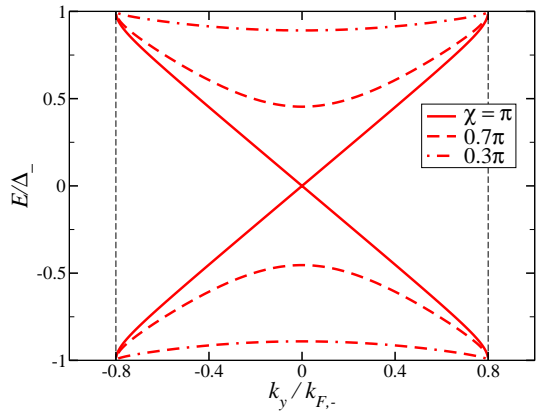}
\caption{(Color online) The surface ABS dispersion in the case of $s$-wave pairing, for $\Delta_+/\Delta_-=1.0$. The critical value of the phase difference is $\chi_c=0$.}
\label{fig: gamma-1.0}
\end{figure}

Let us first consider the TR invariant states. For $\chi=0$, Eq. (\ref{ABS-energy-chi}) does not have any solutions, because 
its left-hand side is negative, while the right-hand side is positive. In contrast, for $\chi=\pi$ there are subgap ABSs, whose energy vanishes at $k_y=0$ according to
\begin{equation}
  E(k_y\to 0)=\pm\frac{k_{F,-}+k_{F,+}}{k_{F,-}k_{F,+}}\frac{\Delta_-\Delta_+}{\Delta_-+\Delta_+}|k_y|,
\end{equation}
see Ref. \onlinecite{TYBN09}.
This can be viewed as a pair of counterpropagating modes with linear dispersion, see the solid lines in Figs. \ref{fig: gamma-0.1} and \ref{fig: gamma-1.0}.
From the topological point of view, the TR invariant states in 2D can be classified by a $Z_2$ invariant, which is equal to the parity of the number of such pairs.\cite{top-SC} 
Thus we have reproduced the known result that the $\chi=0$ state is $Z_2$-trivial and the $\chi=\pi$ state is $Z_2$-nontrivial.\cite{SF09,TYBN09}

If the phase difference $\chi$ is neither $0$ nor $\pi$, then the TR symmetry is broken in the superconducting state and the $Z_2$ topological classification is no longer applicable. 
The ABS energy equation (\ref{ABS-energy-chi}) can be transformed into a biquadratic equation for $E$, supplemented with the
constraint $\Delta_-\Delta_+\cos\chi<E^2<\min(\Delta_-^2,\Delta_+^2)$ [the first inequality makes sure that the left-hand side of Eq. (\ref{ABS-energy-chi}) is positive]. The solution that satisfies the constraint 
has the following form:
\begin{equation}
\label{ABS-energy-solutions-isotropic}
  E(k_y)=\pm\sqrt{\Delta_-\Delta_+F(k_y)},
\end{equation}
where
\begin{eqnarray*}
  F &=& \frac{1}{2(R^2-1)}\biggl[r_+R^2-2\cos\chi\\
  && -R\sqrt{r_-^2R^2+4(1-r_+\cos\chi+\cos^2\chi)}\biggr],
\end{eqnarray*}
with $r_\pm=(\Delta_-^2\pm\Delta_+^2)/\Delta_-\Delta_+$. 
In Figs. \ref{fig: gamma-0.1} and \ref{fig: gamma-1.0}, the ABS dispersion curves are plotted for different values of $\chi$. In all plots we used the same
ratio of the Fermi surface radii, $\rho=0.8$, as shown by the vertical dashed lines. Since the expression (\ref{ABS-energy-solutions-isotropic}) is invariant under the exchange of the gap magnitudes $\Delta_-\leftrightarrow\Delta_+$, 
one can assume that $\Delta_+\leq\Delta_-$.

We see that the Andreev surface modes survive the TR symmetry breaking, but become gapped. The minimum of Eq. (\ref{ABS-energy-solutions-isotropic}) corresponds to $k_y=0$, so that the excitation gap is given by
$$
  E_{gap}\equiv|E(k_y=0)|=\sqrt{\Delta_-\Delta_+}\sqrt{\frac{1-\cos^2\chi}{r_+-2\cos\chi}}.
$$
The gap is zero at $\chi=\pi$ and increases as $\chi$ decreases. One can check that the ABSs exist only if the phase difference between the bands satisfies the condition
\begin{equation}
\label{critical-chi}
  \chi_c<\chi\leq\pi,
\end{equation}
where 
$$
  \chi_c=\arccos\left[\min\left(\frac{\Delta_-}{\Delta_+},\frac{\Delta_+}{\Delta_-}\right)\right].
$$
At $\chi=\chi_c$ the energy gap becomes equal to the lesser of $\Delta_-,\Delta_+$, and the ABS merges into the continuum of bulk states.

\section{Chiral D-wave pairing}
\label{sec: chiral d-wave}

Zero-energy fermionic boundary modes signalling a topologically nontrivial state can also exist in an anisotropically paired superconductor or superfluid with a nonzero phase winding of the gap functions
around the Fermi surface. Such states necessarily break 
TR symmetry, an archetypal example being the chiral $p$-wave state, which is realized in Sr$_2$RuO$_4$ (Ref. \onlinecite{SrRuO}) and in thin films of superfluid ${}^3$He-$A$ (Ref. \onlinecite{VY89}).
In a 2D noncentrosymmetric superconductor, the gap functions with the phase winding can be written as $\tilde\Delta_\lambda(\bk)=\Delta_\lambda e^{i\tilde N\varphi_{\bk}}$.
The phase winding number $\tilde N$ has to be even, because of the condition (\ref{Delta-even}). We focus on the lowest nontrivial case of $\tilde N=2$, which corresponds to the chiral $d$-wave (or $d+id$) 
state of the form $k_x^2-k_y^2+2ik_xk_y$. We further assume that the gap magnitudes in both helicity bands are equal, i.e. $\tilde\Delta_-(\bk)=\tilde\Delta_+(\bk)=\Delta_0e^{2i\varphi_{\bk}}$, with $\Delta_0>0$.

\subsection{$k_{F,+}<|k_y|<k_{F,-}$}
\label{sec: d-wave-N-1}

In this momentum range, we have $N=1$, see Fig. \ref{fig: Rashba-isotropic-N-1}, and the ABS energy equation (\ref{N-1-ABS-energy}) takes the form 
$$
  \frac{E+i\Omega}{E-i\Omega}=e^{-4i\theta_-},\quad\Omega=\sqrt{\Delta_0^2-E^2}.
$$
Its solution is given by $E=-\Delta_0\cos(2\theta_-)\sgn\sin(2\theta_-)$, which can be represented, using Eq. (\ref{ky-theta}), in terms of the momentum parallel to the surface as follows:
\begin{equation}
\label{ABS-energy-anisotropic-1}
  E_1(k_y)=\Delta_0\left(\frac{2k_y^2}{k_{F,-}^2}-1\right)\sgn(k_y).
\end{equation}
The subscript in the energy function signifies the number of the surface scattering channels. We see that the spectrum is odd in $k_y$, $E_1(k_y)=-E_1(-k_y)$, and that there are two zero-energy modes at $k_y=\pm k_{F,-}/\sqrt{2}$, 
which propagate in the same direction. These zeros are located inside the momentum range $k_{F,+}<|k_y|<k_{F,-}$ only if $\rho<1/\sqrt{2}$, see Eq. (\ref{rho-def}), i.e. if the minority Fermi surface is sufficiently small. 
In the single-band limit, when the SO coupling is very strong and $k_{F,+}\to 0$, the results of Ref. \onlinecite{Sam15-1} are recovered.

\subsection{$|k_y|<k_{F,+}$}
\label{sec: d-wave-N-2}

In this momentum range, we have $N=2$, see Fig. \ref{fig: Rashba-isotropic-N-2}, and the ABS energy can be found from Eq. (\ref{N-2-ABS-energy-Rashba}), with $\alpha^{\mathrm{in}}_\pm$ and $\alpha^{\mathrm{out}}_\pm$ 
defined by Eqs. (\ref{alpha-in-def}) and (\ref{alpha-out-def}), respectively. 
It is convenient to introduce the following parametrization: $E=\Delta_0\cos\Theta$, then $\Omega=\Delta_0\sin\Theta\geq 0$.
In terms of $\Theta$, the expressions (\ref{alpha-in-def}) and (\ref{alpha-out-def}) take the form
$\alpha^{\mathrm{in}}_\lambda=(\alpha^{\mathrm{out}}_\lambda)^*=e^{-i(2\theta_\lambda+\Theta)}$. After some straightforward algebra, we arrive at the following equation for $\Theta(k_y)$:
\begin{equation}
\label{E-equation-Rashba-anisotropic}
  \cos(2\theta_-+2\theta_++2\Theta)=P,
\end{equation}
where 
$$
  P(k_y)=1-2\zeta(k_y)\sin^2(\theta_--\theta_+)
$$
and $\zeta$ is given by Eq. (\ref{zeta}). One can easily show that the ABS energy has to be an odd function of momentum. Indeed, since $P$ is even in $k_y$ and $\theta_\pm$ are odd, 
we have $\Theta(-k_y)=-\Theta(k_y)+\pi n$ ($n$ is an integer), and
$$
  E(-k_y)=(-1)^nE(k_y),\quad \Omega(-k_y)=(-1)^{n+1}\Omega(k_y).
$$
It follows from the second of these expressions that $n$ has to be odd, therefore $E(-k_y)=-E(k_y)$.

Focusing on $k_y\geq 0$, we obtain two solutions of Eq. (\ref{E-equation-Rashba-anisotropic}):
\begin{equation}
\label{ABS-energy-anisotropic-21}
  E_2^{(1)}(k_y)=-\Delta_0\cos\left(\theta_-+\theta_+-\frac{1}{2}\arccos P\right)
\end{equation}
and
\begin{equation}
\label{ABS-energy-anisotropic-22}
  E_2^{(2)}(k_y)=-\Delta_0\cos\left(\theta_-+\theta_++\frac{1}{2}\arccos P\right),
\end{equation}
where the reflection angles can be expressed in terms of the momentum parallel to the surface using Eq. (\ref{ky-theta}) and the subscripts in the energy functions signify the number of the surface scattering channels. 
The corresponding ABS dispersion curves are shown in Figs.  \ref{fig: Rashba-anisotropic-r-0.8}, \ref{fig: Rashba-anisotropic-r-0.5}, and \ref{fig: Rashba-anisotropic-r-0.2}.
The $E_2^{(1)}$ branch (shown in red) varies between $-\Delta_0$ at $k_y=+0$ and $(2\rho^2-1)\Delta_0$ at $k_y=k_{F,+}$, where, according to Eq. (\ref{ABS-energy-anisotropic-1}), it connects with the $E_1$ branch.\cite{Andreev-fail} 
The $E_2^{(2)}$ branch (shown in blue) varies between the bulk gap edges, $-\Delta_0$ at $k_y=+0$ and $\Delta_0$ at $k_y=k_{F,+}$, passing through zero in between.

\subsection{Summary}
\label{sec: d-wave-summary}

\begin{figure}
\includegraphics[width=7cm]{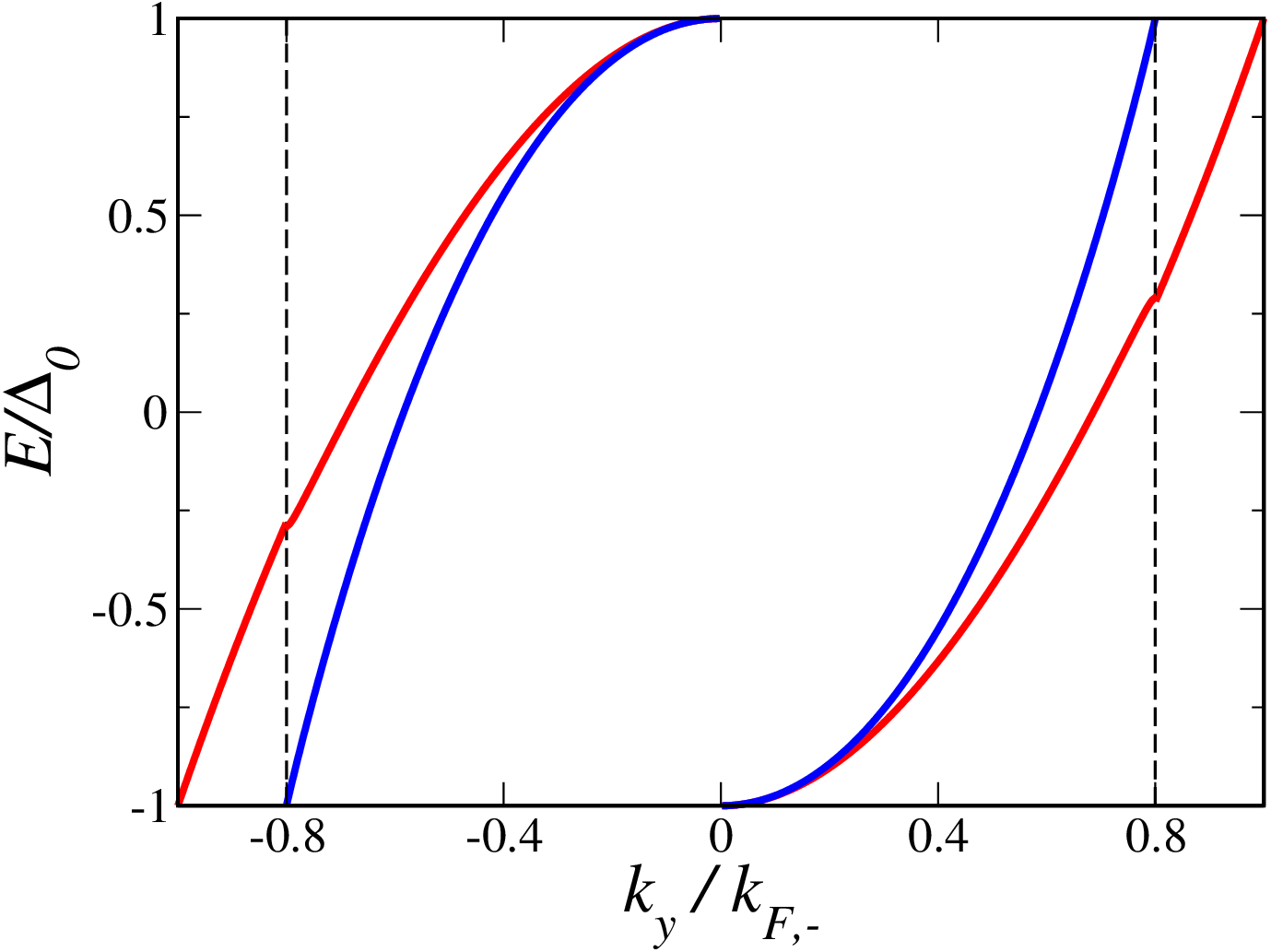}
\caption{(Color online) The surface ABS dispersion in the chiral $d+id$ state, for $\rho=k_{F,+}/k_{F,-}=0.8$. The red and blue lines correspond to different nondegenerate spectral branches, see the text.}
\label{fig: Rashba-anisotropic-r-0.8}
\end{figure}

\begin{figure}
\includegraphics[width=7cm]{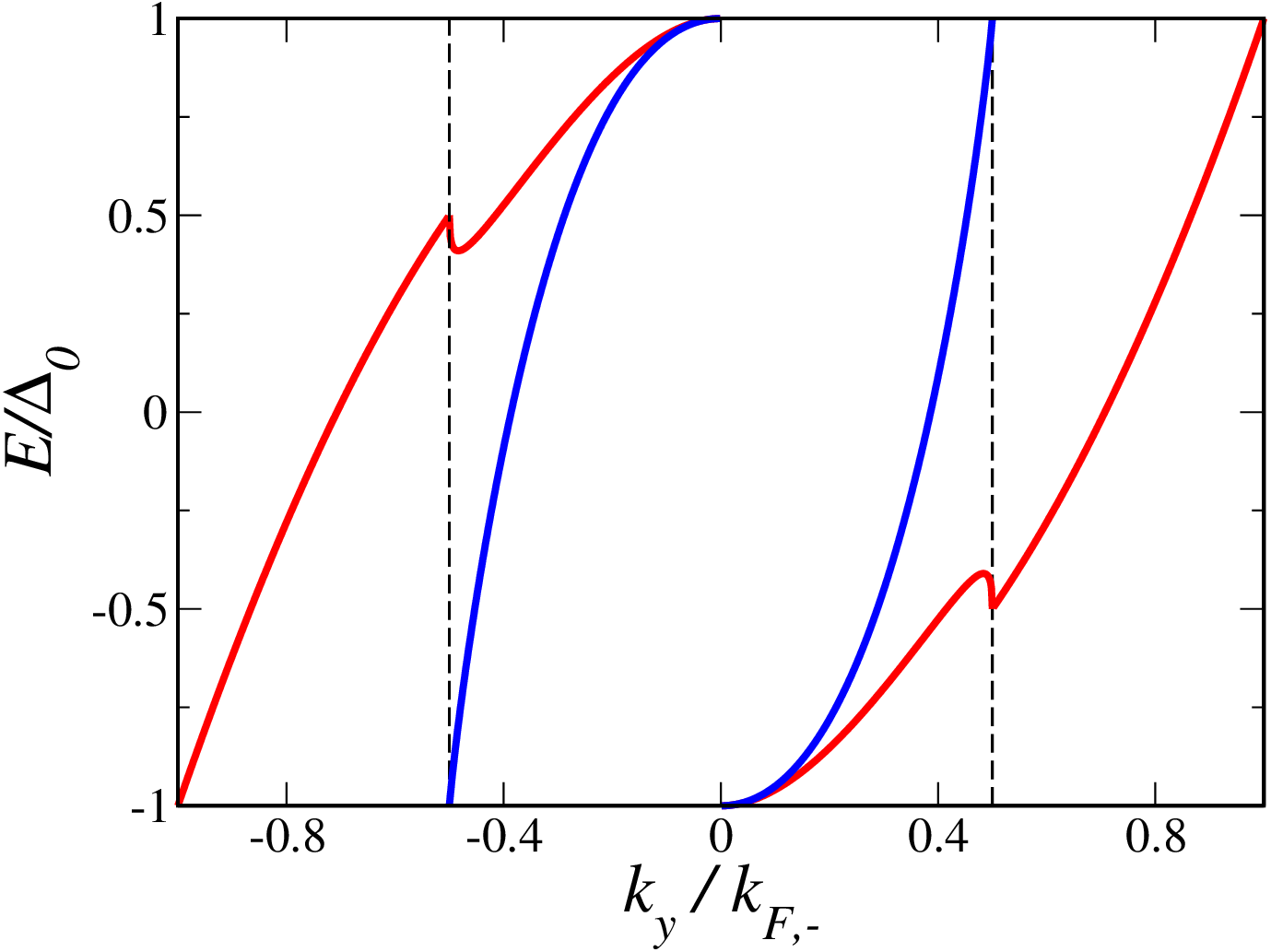}
\caption{(Color online) The surface ABS dispersion in the chiral $d+id$ state, for $\rho=k_{F,+}/k_{F,-}=0.5$. The red and blue lines correspond to different nondegenerate spectral branches, see the text.}
\label{fig: Rashba-anisotropic-r-0.5}
\end{figure}

\begin{figure}
\includegraphics[width=7cm]{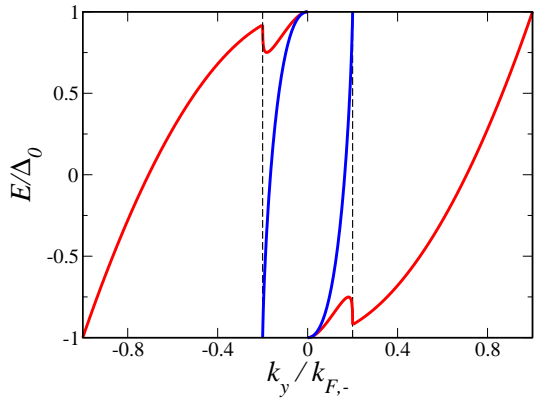}
\caption{(Color online) The surface ABS dispersion in the chiral $d+id$ state, for $\rho=k_{F,+}/k_{F,-}=0.2$. The red and blue lines correspond to different nondegenerate spectral branches, see the text.}
\label{fig: Rashba-anisotropic-r-0.2}
\end{figure}

Numerical investigation of the solutions (\ref{ABS-energy-anisotropic-1}), (\ref{ABS-energy-anisotropic-21}), and (\ref{ABS-energy-anisotropic-22}) reveals a picture of the ABS spectrum which essentially depends on the SO coupling strength. The 
latter is characterized by the ratio of the Fermi momenta $k_{F,+}$ and $k_{F,-}$, see Eq. (\ref{rho-def}).
For most values of $\rho$, there are four symmetrically located zero-energy modes, which propagate in the same (positive) direction along the $y$ axis, as determined by the slopes of the dispersion functions. 
This is shown in Figs. \ref{fig: Rashba-anisotropic-r-0.8}, \ref{fig: Rashba-anisotropic-r-0.5}, and \ref{fig: Rashba-anisotropic-r-0.2}, for three different values of $\rho$. The blue lines correspond to the $E_2^{(2)}$ branch,
while the red lines denote both the $E_2^{(1)}$ branch, at $|k_y|<k_{F,+}$, and the $E_1$ branch, at $k_{F,+}<|k_y|<k_{F,-}$. The vertical dashed lines at $|k_y|=k_{F,+}=\rho k_{F,-}$ show the size of the minority Fermi surface.
The ABS energy is odd in $k_y$ and has a discontinuity at $k_y=0$. The latter is due to the fact that there is no ABS for the normal incidence, when quasiparticles of both helicities sense the same gap function $\Delta_0$ 
before and after the surface reflection.

\begin{figure}
\includegraphics[width=7.0cm]{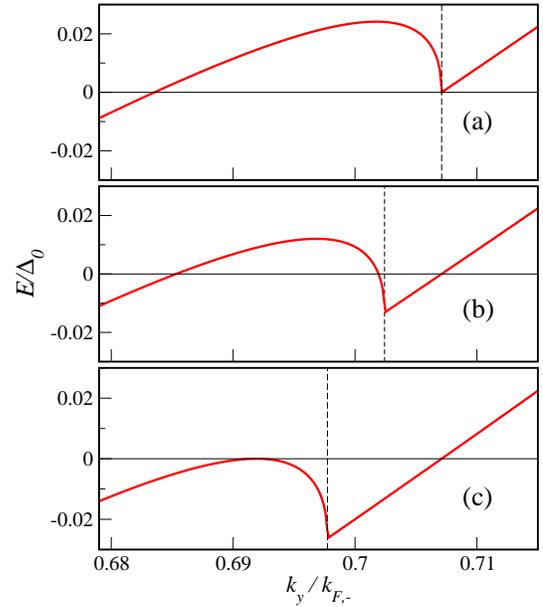}
\caption{(Color online) Evolution of the zero modes in the $E_2^{(1)}$ and $E_1$ branches at $\rho$ close to $1/\sqrt{2}$. The curves (a), (b), and (c) correspond to $\rho=1/\sqrt{2}$, $1/\sqrt{2}-0.5\delta$, 
and $1/\sqrt{2}-\delta$, respectively ($\delta\simeq 9.37\times 10^{-3}$).}
\label{fig: blow-up}
\end{figure}

The ABS dispersion is not a monotonic function of $k_y$, in general, and its slope is discontinuous at $|k_y|=k_{F,+}$  (Ref. \onlinecite{Andreev-fail}). As a consequence, there exists a narrow window of the values of $\rho$ close to 
$1/\sqrt{2}\simeq 0.71$, in which the total number of the ABS zero modes increases to eight, with three pairs propagating in the positive $y$ direction and one pair -- in the negative $y$ direction. 
As the SO band splitting increases, i.e. $\rho$ decreases and reaches $1/\sqrt{2}$, two extra zeros appear at $|k_y|=k_{F,+}$.  These zeros first move apart and then, at $\rho=1/\sqrt{2}-\delta$ (numerically, $\delta\simeq 9.37\times 10^{-3}$), 
one of them merges with and ``cancels'' the other zero mode in the $E_2^{(1)}$ branch. This behaviour is shown in Fig. \ref{fig: blow-up}. 
Both the emergence of the additional zero modes and the numerical smallness of $\delta$ are rather surprising, being most likely artifacts of the isotropic Rashba model. 

In the limit of vanishing SO band splitting, we have $k_{F,-}=k_{F,+}=k_F$. 
The $E_1$ branch disappears and the $E_2$ branches merge, producing two pairs of degenerate zero-energy modes at $k_y=\pm k_F/\sqrt{2}$. Thus we recover the result of Ref. \onlinecite{Volovik97} for the chiral $d$-wave state in a superconductor
without SO coupling.

\subsection{Topological analysis}
\label{sec: chiral dwave-topology}

For all values of $\rho$, the difference between the number of ABS modes moving in the positive $y$ direction and the number of modes moving in the negative $y$ direction is equal to four,
which is a manifestation of the bulk-boundary correspondence.\cite{Volovik-book} The latter stipulates that 
the number of zero-energy surface modes is determined by a topological invariant characterizing the superconducting state in the bulk. To identify the topological invariant appropriate for the chiral $d$-wave state, 
we begin by introducing the Bogoliubov-de Gennes (BdG) Hamiltonian associated with Eq. (\ref{H_mean-field}):
\begin{equation}
\label{H-BdG}
{\cal H}_{BdG}(\bk)=\sum_{\lambda=\pm}\hat\Pi_\lambda(\bk)\otimes\hat h_\lambda(\bk),
\end{equation}
where $\hat\Pi_\lambda(\bk)=|\bk,\lambda\rangle\langle\bk,\lambda|$ is the projector onto the $\lambda$th helicity band,
\begin{equation}
\label{h-def}
\hat h_\lambda(\bk)=\left(\begin{array}{cc}
\xi_\lambda(\bk) & \Delta_\lambda(\bk) \\
\Delta_\lambda^*(\bk) & -\xi_\lambda(\bk)
\end{array}\right)=\bm{\nu}_\lambda(\bk)\hat{\bm{\tau}},
\end{equation}
and
$$
\bm{\nu}_\lambda(\bk)=\left(\begin{array}{c}
\re\Delta_\lambda(\bk)\\
-\im\Delta_\lambda(\bk)\\
\xi_\lambda(\bk)
\end{array}\right).
$$
The BdG Hamiltonian is represented by a $4\times 4$ matrix in the helicity $\times$ electron-hole (Nambu) space, and $\hat{\bm{\tau}}$ are the Pauli matrices in the Nambu space. It is easy to see that $\hat\tau_2\hat h_\lambda(\bk)\hat\tau_2=-\hat h^*_\lambda(\bk)$, which leads to the electron-hole symmetry of the spectrum:
the eigenstates of ${\cal H}_{BdG}(\bk)$ come in pairs, $\pm E_\lambda(\bk)$, where
$$
	E_\lambda(\bk)=|\bm{\nu}_\lambda(\bk)|=\sqrt{\xi_\lambda^2(\bk)+|\Delta_\lambda(\bk)|^2}
$$ 
is the energy of the Bogoliubov excitations in the $\lambda$th band.

Next, we introduce an auxiliary real variable $k_0$ (``frequency'') and 
define the BdG Green's function as follows: ${\cal G}(\bk,k_0)=[ik_0-{\cal H}_{BdG}(\bk)]^{-1}$. 
The topological invariant is constructed in the following way:\cite{Stone-book,Volovik-book}
\begin{equation}
\label{MC-I3}
N_{2+1}=-\frac{1}{24\pi^2}\int\Tr({\cal G}d{\cal G}^{-1})^3,
\end{equation}
where ``$\Tr$'' stands for $4\times 4$ matrix trace and combined matrix and exterior multiplication is implied inside the trace. The integration is performed over a closed $(2+1)$-dimensional manifold with coordinates $k_x,k_y,k_0$, 
which is topologically equivalent to a 3D torus (the frequency variable runs over the real axis, which is assumed to be closed into a circle). Calculating the trace and integrating over $k_0$, we arrive at 
the following expression:\cite{Sam15-1}
\begin{equation}
\label{N3-nu}
N_{2+1}=\frac{1}{8\pi}\sum_\lambda\int_{BZ}\hat{\bm{\nu}}_\lambda(d\hat{\bm{\nu}}_\lambda\times d\hat{\bm{\nu}}_\lambda),
\end{equation} 
where $\hat{\bm{\nu}}_\lambda=\bm{\nu}_\lambda/|\bm{\nu}_\lambda|$.
Note that the integrand is nonzero only inside the BCS momentum shells near the Fermi surfaces, since 
$\hat{\bm{\nu}}_\lambda=\hat{\bm z}\sgn\xi_\lambda(\bk)$ outside the BCS shells.

Writing the gap functions in the form
\begin{equation}
 \label{nodeless gap}
 \Delta_\lambda(\bk)=|\Delta_\lambda(\bk)|e^{i\Phi_\lambda(\bk)},
\end{equation}
assuming a fully gapped superconducting state, and integrating over $\xi_\lambda$, we finally obtain:
\begin{equation}
\label{total-winding-number}
N_{2+1}=\sum_\lambda N_\lambda,
\end{equation}
where
$$
N_\lambda=\frac{1}{2\pi}\oint_{FS_\lambda}d\Phi_\lambda
$$
is the winding number of the gap phase $\Phi_\lambda(\bk)$ along the $\lambda$th Fermi surface. For the chiral $d$-wave state considered here, we have $N_-=N_+=2$ and $N_{2+1}=4$.

\section{Nonchiral D-wave pairing}
\label{sec: d-wave}

In this section we consider the TR invariant $d$-wave states of the form $\tilde\Delta_\lambda(\bk)\propto k_xk_y$ or $k_x^2-k_y^2$, referred to as $d_{xy}$ or $d_{x^2-y^2}$ states, respectively. 
For simplicity, we assume the same gap magnitudes in both helicity bands. Then, we have 
\begin{equation}
\label{dxy}
  \tilde\Delta_-(\bk)=\tilde\Delta_+(\bk)=\Delta_0\sin(2\varphi_{\bk})
\end{equation}
or
\begin{equation}
\label{dx2y2}
  \tilde\Delta_-(\bk)=\tilde\Delta_+(\bk)=\Delta_0\cos(2\varphi_{\bk}),
\end{equation}
where $\Delta_0>0$ and $\varphi_{\bk}=\tan^{-1}(k_y/k_x)$ is the angle between $\bk$ and the positive $x$ axis.

\subsection{$d_{xy}$ state}
\label{sec: dxy}

Generally, the ABS formation is only possible when the gap function is not constant along the quasiparticle's trajectory.
As seen from Figs. (\ref{fig: Rashba-isotropic-N-1}) and (\ref{fig: Rashba-isotropic-N-2}), the gap functions (\ref{dxy}) have opposite signs on the incident and reflected trajectories, regardless of the helicity and the value of $k_y$: 
\begin{equation}
\label{dxy-Deltas}
  \Delta(\bk^{\mathrm{out}}_\lambda)=-\Delta(\bk^{\mathrm{in}}_\lambda)=\Delta_0\sin(2\theta_\lambda)\equiv\Delta_\lambda(k_y).
\end{equation}
Therefore, one can expect the presence of the ABS zero-energy modes, similar to those in centrosymmetric $d$-wave superconductors.\cite{Hu94,ZBCP} Below this is confirmed by a direct calculation. 

\begin{figure}
\includegraphics[width=7cm]{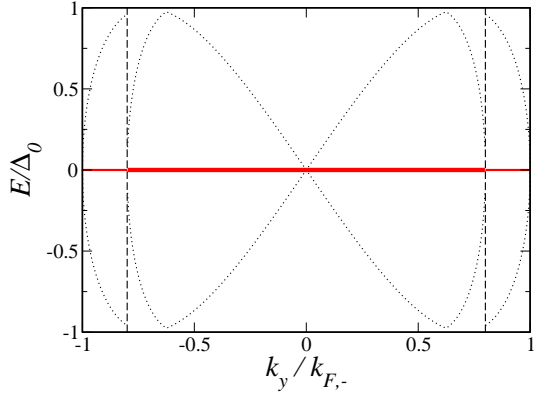}
\caption{(Color online) The surface ABS dispersion in the $d_{xy}$ state, for $\rho=0.8$. The vertical dashed lines at $k_y=\pm\rho k_{F,-}$ show the size of the minority Fermi surface. 
The dotted lines denote the bulk gap edge $\Delta_b(k_y)$. The zero-energy states are doubly degenerate at $|k_y|<k_{F,+}$
and nondegenerate at $k_{F,+}<|k_y|<k_{F,-}$, see Eq. (\ref{N1-dxy}). }
\label{fig: dxy}
\end{figure}

At $k_{F,+}<|k_y|<k_{F,-}$, we obtain from Eq. (\ref{N-1-ABS-energy}):
\begin{equation}
\label{dxy-N-1}
  \frac{E+i\sqrt{\Delta_-^2-E^2}}{E-i\sqrt{\Delta_-^2-E^2}}=-1.
\end{equation}
At $|k_y|<k_{F,+}$, the ABS energy equation has the form (\ref{N-2-ABS-energy-Rashba}), which can be reduced to
\begin{equation}
\label{dxy-N-2}
  \frac{\sqrt{(\Delta_-^2-E^2)(\Delta_+^2-E^2)}-\Delta_-\Delta_+}{E^2}=R,
\end{equation}
where $R(k_y)$ is the same as in Eq. (\ref{R-ky}). It follows from Eq. (\ref{dxy-Deltas}) that $\Delta_-(k_y)\Delta_+(k_y)\geq 0$, from which one concludes that the only solution of Eqs. (\ref{dxy-N-1}) and (\ref{dxy-N-2}) is
\begin{equation}
\label{dxy-ABS-energy}
  E(k_y)=0,
\end{equation}
at all $k_y$, regardless of the SO coupling strength. Thus we have reproduced the dispersionless ABS spectrum in the $d_{xy}$ state found previously in Ref. \onlinecite{flat-ABS}.
The ABS energy is shown by the solid red line in Fig. \ref{fig: dxy}, along with the anisotropic bulk gap edge $\Delta_b(k_y)$. The latter is given by 
\begin{equation}
\label{Delta-b-N-1}
  \Delta_b(k_y)=|\Delta_-(k_y)|
\end{equation}
at $k_{F,+}<|k_y|<k_{F,-}$, and
\begin{equation}
\label{Delta-b-N-2}
  \Delta_b(k_y)=\min\{|\Delta_-(k_y)|,|\Delta_+(k_y)|\}
\end{equation}
at $|k_y|<k_{F,+}$. These zero-energy states have a topological origin, as shown in Ref. \onlinecite{SchRyu} 
and also discussed in Sec. \ref{sec: nonchiral d-wave topology} below.

\subsection{$d_{x^2-y^2}$ state}
\label{sec: dx2y2}

The expressions (\ref{dx2y2}) take the following values on the incident and reflected trajectories:  
\begin{equation}
\label{dx2y2-Deltas}
  \Delta(\bk^{\mathrm{out}}_\lambda)=\Delta(\bk^{\mathrm{in}}_\lambda)=\Delta_0\cos(2\theta_\lambda)\equiv\Delta_\lambda(k_y).
\end{equation}
The gap function remains unchanged after the surface reflection into the same helicity band, therefore there are no subgap ABSs for $N=1$, i.e. at $k_{F,+}<|k_y|<k_{F,-}$. 
However, if the quasiparticle is reflected into the opposite-helicity band at $N=2$, then the gap functions before and after the surface reflection can differ, leading to the possibility of a subgap ABS at $|k_y|<k_{F,+}$.

\begin{figure}
\includegraphics[width=7cm]{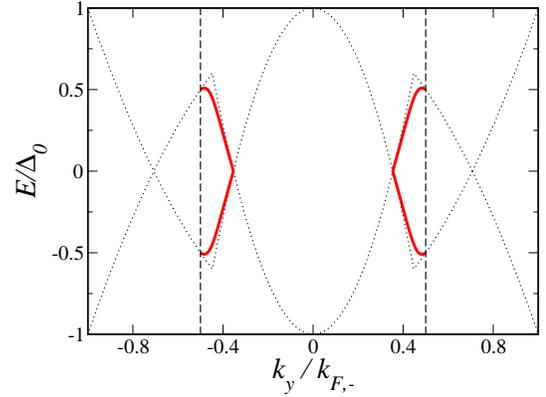}
\caption{(Color online) The surface ABS dispersion in the $d_{x^2-y^2}$ state, for $\rho=0.5$. The vertical dashed lines at $k_y=\pm\rho k_{F,-}$ show the size of the minority Fermi surface. 
The dotted lines denote the bulk gap edge $\Delta_b(k_y)$.}
\label{fig: dx2y2-r05}
\end{figure}

\begin{figure}
\includegraphics[width=7cm]{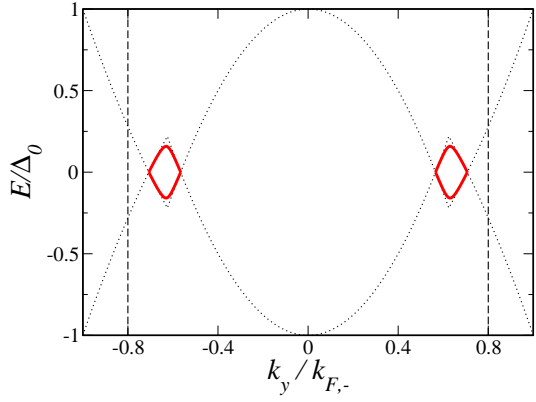}
\caption{(Color online) The surface ABS dispersion in the $d_{x^2-y^2}$ state, for $\rho=0.8$. The vertical dashed lines at $k_y=\pm\rho k_{F,-}$ show the size of the minority Fermi surface. 
The dotted lines denote the bulk gap edge $\Delta_b(k_y)$.}
\label{fig: dx2y2-r08}
\end{figure}

From Eq. (\ref{N-2-ABS-energy-Rashba}) we obtain
\begin{equation}
\label{dx2y2-N-2}
  \frac{E^2-\Delta_-\Delta_+}{\sqrt{(\Delta_-^2-E^2)(\Delta_+^2-E^2)}}=R,
\end{equation}
where $\Delta_\pm(k_y)$ are defined in Eq. (\ref{dx2y2-Deltas}) and $R(k_y)$ is given by Eq. (\ref{R-ky}). The last equation has a subgap solution only if $\Delta_-(k_y)\Delta_+(k_y)<0$, which is realized at
\begin{equation}
\label{k-window}
    \frac{\rho}{\sqrt{2}}<\frac{|k_y|}{k_{F,-}}<\min\left\{\rho,\frac{1}{\sqrt{2}}\right\}.
\end{equation}
In this momentum range, $\Delta_-(k_y)>0$, but $\Delta_+(k_y)<0$, and the ABS energy has the following form:
\begin{equation}
\label{dx2y2-ABS-energy}
  E(k_y)=\pm\sqrt{\Phi(k_y)},
\end{equation}
where
\begin{eqnarray*}
  \Phi &=& \frac{1}{2(R^2-1)}\biggl[R^2(\Delta_-^2+\Delta_+^2)-2\Delta_-\Delta_+\\
    && -R(\Delta_--\Delta_+)\sqrt{R^2(\Delta_-+\Delta_+)^2-4\Delta_-\Delta_+})\biggr].
\end{eqnarray*}
We have plotted the dispersion curves (shown by the solid red lines) in Figs. \ref{fig: dx2y2-r05} and \ref{fig: dx2y2-r08}, along with the anisotropic bulk gap edge $\Delta_b(k_y)$, defined in Eqs. (\ref{Delta-b-N-1}) and (\ref{Delta-b-N-2}).
The subgap ABS modes are present only in a rather narrow window of momenta along the surface, satisfying the condition (\ref{k-window}).

\subsection{Topological analysis}
\label{sec: nonchiral d-wave topology}

The presence of the zero-energy ABSs in the $d_{xy}$ state and their absence in the $d_{x^2-y^2}$ state can also be understood using topological arguments.\cite{SchRyu} 
In a TR invariant superconducting state, the gap functions are real and we have $\hat\tau_2\hat h_\lambda(\bk)\hat\tau_2=-\hat h_\lambda(\bk)$. Therefore, the BdG Hamiltonian (\ref{H-BdG}) has the ``chiral'' symmetry:
\begin{equation}
\label{chiral-symmetry}
  \{{\cal C},{\cal H}_{BdG}(\bk)\}=0,\quad {\cal C}=\hat{\mathbb{1}}\otimes\hat\tau_2.
\end{equation}
Note that $\hat h_\lambda(\bk)=\hat h_\lambda(-\bk)$, due to Eq. (\ref{Delta-even}). 

Superconducting states in the bulk can be classified into different universality classes, according to the topology of the mapping 
$\bk\to{\cal H}_{BdG}(\bk)$. These universality classes are characterized by topological invariants obtained by integrating certain differential forms constructed from
${\cal H}_{BdG}$ over closed manifolds in momentum space. For the TR invariant states, the relevant topological invariant has the following form:
\begin{equation}
\label{MC-invariant}
  N^{\mathrm{TRI}}_1=\frac{1}{4\pi i}\oint\Tr\left({\cal C}{\cal H}^{-1}_{BdG}d{\cal H}_{BdG}\right),
\end{equation}
where the integration is performed over a closed 1D contour in the momentum space. One can show that this last expression remains unchanged under any small variation of the system parameters which respects the symmetry (\ref{chiral-symmetry}),
see, e.g., Ref. \onlinecite{Sam15-1}. 

A straightforward calculation yields 
$$
  {\cal H}^{-1}_{BdG}d{\cal H}_{BdG}=\sum_{\lambda_1\lambda_2}\hat\Pi_{\lambda_1}d\hat\Pi_{\lambda_2}\otimes \hat P_{\lambda_1\lambda_2}+\sum_\lambda\hat\Pi_\lambda\otimes\hat Q_\lambda,
$$
where 
\begin{eqnarray*}
  && \hat P_{\lambda_1\lambda_2}=\frac{E_{\lambda_2}}{E_{\lambda_1}}\left[(\hat{\bm{\nu}}_{\lambda_1}\hat{\bm{\nu}}_{\lambda_2})\hat\tau_0+i(\hat{\bm{\nu}}_{\lambda_1}\times\hat{\bm{\nu}}_{\lambda_2})\hat{\bm{\tau}}\right],\\
  && \hat Q_\lambda=\frac{dE_\lambda}{E_\lambda}\hat\tau_0+i(\hat{\bm{\nu}}_\lambda\times d\hat{\bm{\nu}}_\lambda)\hat{\bm{\tau}}
\end{eqnarray*}
are Nambu matrix-valued $0$- and $1$-forms, respectively. Inserting these expressions in Eq. (\ref{MC-invariant}) and calculating the traces, we obtain:
\begin{eqnarray}
\label{N-TRI-final}
  N^{\mathrm{TRI}}_1&=&\frac{1}{2\pi}\sum_\lambda\oint\frac{\xi_\lambda d\Delta_\lambda-\Delta_\lambda d\xi_\lambda}{E_\lambda^2}\nonumber\\
  &=&\frac{1}{2\pi}\sum_\lambda\oint d\tilde\Phi_\lambda,
\end{eqnarray}
where $\tilde\Phi_\lambda(\bk)$ is the phase of the complex number $\xi_\lambda(\bk)+i\Delta_\lambda(\bk)$. The last integral vanishes unless the integration contour encloses one or more points where $\tilde\Phi_\lambda$ 
is not defined, i.e. the gap nodes on the Fermi surface, where $\xi_\lambda(\bk)=\Delta_\lambda(\bk)=0$.

\begin{figure}
	\includegraphics[width=5.5cm]{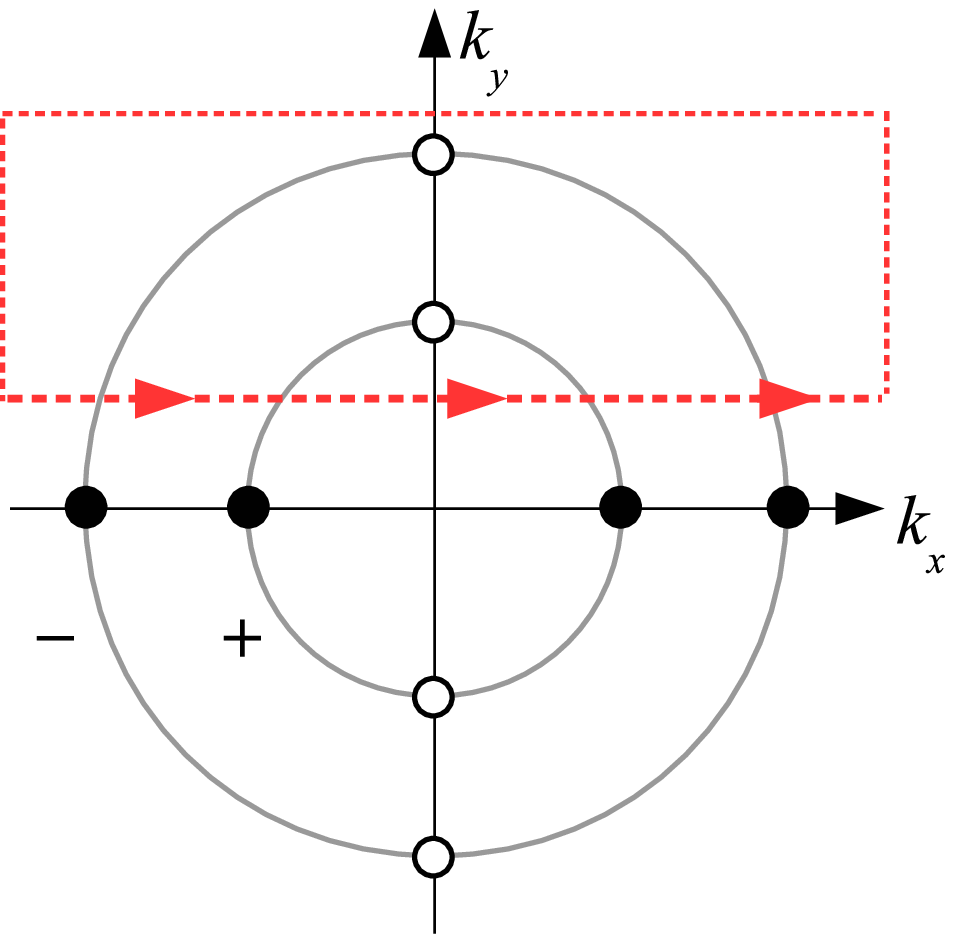}
	\caption{(Color online) The integration contour in Eq. (\ref{N-TRI-final}) for the $d_{xy}$ state. The filled (empty)
		dots correspond to the gap nodes with the topological charge $+1$ ($-1$).}
	\label{fig: dxy-nodes}
\end{figure}

\begin{figure}
	\includegraphics[width=5.5cm]{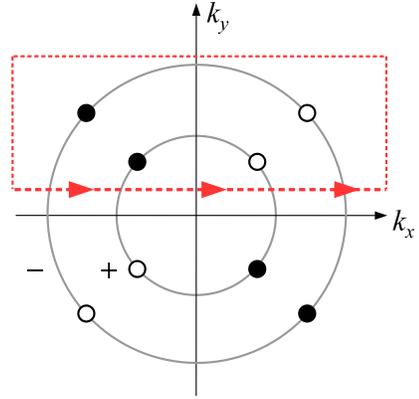}
	\caption{(Color online) The integration contour in Eq. (\ref{N-TRI-final}) for the $d_{x^2-y^2}$ state. The filled (empty)
		dots correspond to the gap nodes with the topological charge $+1$ ($-1$).}
	\label{fig: dx2y2-nodes}
\end{figure}

According to Ref. \onlinecite{SchRyu}, in order to count the zero-energy ABS modes at given momentum along the surface, one should integrate in Eq. (\ref{N-TRI-final}) along 
a straight line running from to $k_x=-\infty$ to $+\infty$ (or between the opposite edges of the BZ). Then the number of the zero-energy ABSs is equal to $|N^{\mathrm{TRI}}_1(k_y)|$. 
One can now use Stokes' theorem to contract the integration contour by deforming it through the BZ without crossing any gap nodes, see Figs. \ref{fig: dxy-nodes} 
and \ref{fig: dx2y2-nodes}, and show that
\begin{equation}
\label{I-1-final}
  N^{\mathrm{TRI}}_1=\sum_{\lambda,i} q_{\lambda,i}.
\end{equation}
Here the sum is taken only over the gap nodes enclosed by the contour, 
\begin{equation}
\label{top-charge}
  q_{\lambda,i}=\frac{1}{2\pi}\oint_{c_{\lambda,i}} d\tilde\Phi_\lambda
\end{equation}
has the meaning of the topological charge of the $i$th gap node in the $\lambda$th band, and $c_{\lambda,i}$ is an infinitesimally small circular contour wrapping counterclockwise around the node. 

The gap nodes in the $d_{xy}$ state 
are located at $k_x=0,|k_y|=k_{F,\lambda}$ and at $k_y=0,|k_x|=k_{F,\lambda}$, while in the 
$d_{x^2-y^2}$ state they are located at $|k_x|=|k_y|=k_{F,\lambda}/\sqrt{2}$. Their topological charges are equal to either $+1$
or $-1$, as shown in Figs. \ref{fig: dxy-nodes} and \ref{fig: dx2y2-nodes} by the filled or empty dots, respectively.  
From Eq. (\ref{I-1-final}) we finally obtain:
\begin{equation}
\label{N1-dxy}	
	|N^{\mathrm{TRI}}_1(k_y)|=\left\{
	\begin{array}{ll}
	2,\quad & \mathrm{at}\ |k_y|<k_{F,+},\medskip\\ 
	1,\quad & \mathrm{at}\ k_{F,+}<|k_y|<k_{F,-},
	\end{array}
	\right.
\end{equation}
in the $d_{xy}$ state, and
\begin{equation}
\label{N1-dx2y2}	
|N^{\mathrm{TRI}}_1(k_y)|=0,\quad \mathrm{at\ all}\ k_y,
\end{equation}
in the $d_{x^2-y^2}$ state. Thus, the zero-energy ABSs in the $d_{xy}$ state are topologically protected,
while one should generally not expect the zero modes in the $d_{x^2-y^2}$ state.

\section{Conclusions}
\label{sec: Conclusions}

We have developed a theory of fermionic boundary modes in 2D superconductors without inversion symmetry, in the presence of a strong SO coupling. Due to the band splitting being much greater than the energy scales associated with
superconductivity, the Cooper pairing occurs only between the time-reversed states of the same helicity. The boundary modes appear as the subgap bound states in the semiclassical, or Andreev, equations for the quasiparticle wave 
function. The boundary conditions for the Andreev equations are expressed in terms of the surface $S$-matrix. The advantage of the $S$-matrix formalism is that it can be extended, at least phenomenologically, to describe 
more complicated band structures and other types of the surface scattering, e.g. non-specular and/or TR symmetry-breaking.

In the helicity band representation, the gap functions $\tilde\Delta_-(\bk)$ and $\tilde\Delta_+(\bk)$ are necessarily even in momentum. We have studied in detail various $s$-wave and $d$-wave pairing states, 
both TR symmetry-breaking and TR invariant, and found qualitatively different ABS spectra. We hope that our results 
will be useful for the identification of the gap symmetry in 2D interface superconductors. The boundary modes contribute to the quasiparticle density of states and can therefore be probed in tunneling experiments, which has been successfully done 
in other unconventional superconductors, for instance, in high-$T_c$ cuprates (Ref. \onlinecite{HTSC-tunneling}) and Sr$_2$RuO$_4$ (Ref. \onlinecite{SRO-tunneling}).

The isotropically gapped $s$-wave state is described by $\tilde\Delta_-=\Delta_-$ and $\tilde\Delta_+=\Delta_+e^{i\chi}$, with the phase difference $0\leq\chi\leq\pi$. 
At $\chi=\pi$, there are two counterpropagating zero-energy ABS modes, which corresponds to a $Z_2$-nontrivial topological class. At $\chi<\pi$, 
the TR symmetry is broken and the ABS spectrum develops a gap. There exists a critical value of the phase difference, $0\leq\chi_c<\pi$, at which the ABSs disappear, merging into the continuum of the bulk states. 

In the TR symmetry-breaking $d$-wave state of the form $\tilde\Delta_\pm\propto k_x^2-k_y^2+2ik_xk_y$, the ABS spectrum consists of four nondegenerate chiral branches, with a nonmonotonic dependence on the momentum parallel to the surface. 
These modes can carry a charge current along the boundary of the 2D superconductor. While the total number of the zero-energy modes depends on the SO band splitting, their algebraic number (which takes into account 
the direction of propagation) is a topological invariant equal to the sum of the gap phase winding numbers in the helicity bands. 

In the TR invariant $d$-wave state $\tilde\Delta_\pm\propto k_xk_y$, the ABS modes of zero energy are present at all momenta along the surface. This can be attributed to the fact that the gap function sensed by a quasiparticle along its 
semiclassical trajectory always changes sign upon the surface reflection. In contrast, in the state $\tilde\Delta_\pm\propto k_x^2-k_y^2$ the ABS ``pockets'' exist only in a certain momentum range, where there is a nonzero probability of 
the gap function changing sign due to the helicity flip during the surface scattering. 

\acknowledgments
This work was supported by a Discovery Grant from the Natural Sciences and Engineering Research Council of Canada.

\appendix

\section{$S$-matrix for the Rashba model}
\label{app: S-matrix}

The surface scattering matrix is an electron-hole scalar and can therefore be calculated in the normal state.\cite{Shel-bc} Let us first consider the case of two scattering channels.
The quasiparticle wave function in the bulk at given $k_y$, satisfying $|k_y|<k_{F,+}$, is a superposition of two incident and two reflected waves:
\begin{equation}
\label{Psi-general-2}
 \Psi(\br)=A_-\langle\br|\bk^{\mathrm{in}}_{-}\rangle+A_+\langle\br|\bk^{\mathrm{in}}_{+}\rangle+
  B_-\langle\br|\bk^{\mathrm{out}}_{-}\rangle+B_+\langle\br|\bk^{\mathrm{out}}_{+}\rangle.
\end{equation}
All four states here are located at the Fermi level: $\xi_\lambda(\bk^{\mathrm{in}}_\lambda)=\xi_\lambda(\bk^{\mathrm{out}}_\lambda)=0$.
Using the eigenstates of the Rashba model, see Eq. (\ref{Rashba-eigenstates}), we have
\begin{eqnarray}
\label{chi-in-out}
  && \langle\br|\bk^{\mathrm{in}}_{\lambda}\rangle=\frac{1}{\sqrt{2|v_{\lambda,x}(\bk^{\mathrm{in}}_{\lambda})|}}\left(\begin{array}{c}
                                       1 \\ i\lambda e^{-i\theta_\lambda}
                                      \end{array}\right)e^{i\bk^{\mathrm{in}}_{\lambda}\br},\nonumber\\
  &&                                    \\
  && \langle\br|\bk^{\mathrm{out}}_{\lambda}\rangle=\frac{1}{\sqrt{2|v_{\lambda,x}(\bk^{\mathrm{out}}_{\lambda})|}}\left(\begin{array}{c}
                                       1 \\ -i\lambda e^{i\theta_\lambda}
                                      \end{array}\right)e^{i\bk^{\mathrm{out}}_{\lambda}\br},\nonumber
\end{eqnarray}
where the angles of reflection $\theta_\pm$ are defined in Fig. \ref{fig: reflection angles}. We use the normalization in which the magnitude of the probability current carried in the $x$ direction by each of 
the plane-wave states (\ref{chi-in-out}) is equal to one. It follows from Eq. (\ref{v-F-pm}) that $|v_{\lambda,x}(\bk^{\mathrm{in}}_{\lambda})|=|v_{\lambda,x}(\bk^{\mathrm{out}}_{\lambda})|=v_F\cos\theta_\lambda$. 
The complex amplitudes $A_\pm$ and $B_\pm$ satisfy the condition
\begin{equation}
\label{current-conservation}
  |A_-|^2+|A_+|^2=|B_-|^2+|B_+|^2,
\end{equation}
which expresses the particle number conservation in terms of the equality of the incident and reflected currents.  

From the microscopic boundary condition for the wave function at an infinitely high wall,
\begin{equation}
\label{boundary-condition}
  \Psi(x=0,y)=0,
\end{equation}
we obtain two linear relations between the four coefficients $A_\pm$ and $B_\pm$. These relations can be written in the matrix form as follows: 
$$
  \left(\begin{array}{c}
  B_- \\ B_+
  \end{array}\right)=
  \hat S
  \left(\begin{array}{c}
  A_- \\ A_+
  \end{array}\right),
$$
where 
\begin{eqnarray}
\label{S-matrix-N-2}
  \hat S &=&-\frac{1}{e^{i\theta_-}+e^{i\theta_+}}\nonumber\\
	 && \times\left(\begin{array}{cc}
         e^{i\theta_+}-e^{-i\theta_-} & 2\sqrt{\cos\theta_-\cos\theta_+} \\
         2\sqrt{\cos\theta_-\cos\theta_+} & e^{i\theta_-}-e^{-i\theta_+}
         \end{array}\right),
\end{eqnarray}
see also Ref. \onlinecite{KS10}. The diagonal and off-diagonal elements of the $S$-matrix are related to the probabilities of different surface scattering processes. For the helicity-conserving transitions 
$\bk^{\mathrm{in}}_-\to\bk^{\mathrm{out}}_-$ and $\bk^{\mathrm{in}}_+\to\bk^{\mathrm{out}}_+$, the probabilities are given by $|S_{--}|^2$ and $|S_{++}|^2$, respectively, while for the helicity-flip 
transitions $\bk^{\mathrm{in}}_-\to\bk^{\mathrm{out}}_+$ and $\bk^{\mathrm{in}}_+\to\bk^{\mathrm{out}}_-$ the probability is given by $|S_{-+}|^2=|S_{+-}|^2$. 
The $S$-matrix is unitary, in agreement with the particle number conservation condition (\ref{current-conservation}). As shown in Appendix \ref{app: TRS}, it also satisfies an additional constraint 
imposed by TR invariance, which relates $\hat S(k_y)$ and $\hat S(-k_y)$.

In the case of normal incidence, when $\theta_-=\theta_+-=0$, the $S$-matrix takes a particularly simple form $\hat S=-\hat\sigma_x$. The absence of the diagonal matrix elements can be easily understood: at 
$k_y=0$ the direction of momentum is reversed upon reflection, but the spin is unchanged, which means that the normal scattering flips the sign of helicity. The phase shift of $\pi$ between the incident and reflected waves 
makes sure that the wave function vanishes at the surface.

The case of one scattering channel is realized at $k_{F,+}<|k_y|<k_{F,-}$, when the waves corresponding to the minority band become evanescent in the bulk. 
Although the positive helicity states do not participate in the superconducting pairing, one has to take them 
into account when calculating the normal-state surface scattering matrix, in order to satisfy the boundary condition. The quasiparticle wave function at given $k_y$, with the energy at the Fermi level, now has the form
\begin{equation}
\label{Psi-general-1}
 \Psi(\br)=A_-\langle\br|\bk^{\mathrm{in}}_{-}\rangle+B_-\langle\br|\bk^{\mathrm{out}}_{-}\rangle+\tilde\psi_+(\br).
\end{equation}
The first two terms are the propagating wave states in the majority band, see Eq. (\ref{chi-in-out}), and the last term is the minority-band evanescent state given by
$$
  \tilde\psi_+(\br)=C\left(\begin{array}{c}
               \dfrac{k_y-\kappa}{k_{F,+}} \\ 1
               \end{array}\right)e^{-\kappa x}e^{ik_yy},
$$
where $\kappa=\sqrt{k_y^2-k_{F,+}^2}$ and $C$ is a coefficient. From the boundary condition (\ref{boundary-condition}) we obtain the following expression for the only element of the $S$-matrix:
\begin{equation}
\label{S-matrix-N-1}
  S_{--}=\frac{B_-}{A_-}=-\frac{k_{F,+}+i(k_y-\kappa)e^{-i\theta_-}}{k_{F,+}-i(k_y-\kappa)e^{i\theta_-}}.
\end{equation}
It is easy to see that $|S_{--}|=1$, in agreement with the particle number conservation, which requires $|B_-|^2=|A_-|^2$.

\section{TR symmetry of the $S$-matrix}
\label{app: TRS}

We assume that there are $N$ surface scattering channels and introduce the shorthand notations $|\sigma\rangle\equiv|\bk,\lambda\rangle$ and $|\bar\sigma\rangle\equiv|-\bk,\lambda\rangle$.
These two states have the same energy and are connected by the time reversal operation: $K|\sigma\rangle=t(\sigma)|\bar\sigma\rangle$,
where $t(\sigma)\equiv t_\lambda(\bk)=-t_\lambda(-\bk)=-t(\bar\sigma)$ is a phase factor, see Eq. (\ref{t-lambda}). The general wave function in the bulk has the following form, cf. Eq. (\ref{Psi-general-2}):
\begin{equation}
\label{Psi-general-N}
 |\Psi\rangle=\sum_{i=1}^N\left(A_i|\sigma_i\rangle+B_i|\sigma'_i\rangle\right),
\end{equation}
where the $A$s are the amplitudes of the incident states $|\sigma\rangle$ and the $B$s the amplitudes of the reflected states $|\sigma'\rangle$. 
The surface scattering matrix is defined by the equations
\begin{equation}
\label{S-matrix-N}
  B_i=\sum_{j=1}^NS(\sigma'_i,\sigma_j)A_j.
\end{equation}
Applying the TR operation to the wave function (\ref{Psi-general-N}), we obtain 
\begin{equation}
\label{K-Psi-general-N}
 K|\Psi\rangle=\sum_{i=1}^N\left[A_i^*t(\sigma_i)|\bar\sigma_i\rangle+B_i^*t(\sigma'_i)|\bar\sigma'_i\rangle\right].
\end{equation}
Here the states $|\bar\sigma\rangle$ correspond to reflected waves, while the states $|\bar\sigma'\rangle$ correspond to incident waves. 

If the bulk Hamiltonian and the surface scattering are both TR invariant, then one can expect the same 
$S$-matrix relations between the incident and reflected states in $|\Psi\rangle$ and $K|\Psi\rangle$, therefore
\begin{equation}
\label{TR-S-matrix-N}
  A_i^*t(\sigma_i)=\sum_{j=1}^NS(\bar\sigma_i,\bar\sigma'_j)B_j^*t(\sigma'_j).
\end{equation}
Comparing Eqs. (\ref{S-matrix-N}) and (\ref{TR-S-matrix-N}) and taking into account the unitarity of the $S$-matrix, expressed as
$$
  \sum_{k=1}^N S(\sigma'_i,\sigma_k)S^*(\sigma'_j,\sigma_k)=\delta_{ij}, 
$$
we arrive at the following constraints imposed by TR symmetry:
$S(\bar\sigma_j,\bar\sigma'_i)=t^*(\sigma'_i)S(\sigma'_i,\sigma_j)t(\sigma_j)$, or, more explicitly,
$$
  S_{\lambda\lambda'}(-\bk,-\bk')=t^*_{\lambda'}(\bk')S_{\lambda'\lambda}(\bk',\bk)t_{\lambda}(\bk).
$$
In particular, in the Rashba model the phase factor is given by $t_\lambda(\bk)=i\lambda e^{-i\varphi_{\bk}}$ and, if the momentum along the surface is conserved, we obtain: 
$$
  S_{\lambda\lambda'}(-k_y)=-\lambda\lambda'e^{i(\theta_\lambda+\theta_{\lambda'})}S_{\lambda'\lambda}(k_y).
$$
It is straightforward to check that the $S$-matrices (\ref{S-matrix-N-2}) and (\ref{S-matrix-N-1}) satisfy this last condition.


\begin{thebibliography}{999}
\bibitem{Andreev-fail}
The Andreev approximation is not applicable at $|k_y|=k_{F,+}$, because the corresponding trajectories for the quasiparticles in the 
$\lambda=+$ band are parallel to the surface. For the same reason, it is not applicable at $|k_y|=k_{F,-}$.
\end{thebibliography}
\end{document}